 \documentclass[final,5p,times,twocolumn]{elsarticle}

\usepackage{amssymb}
\usepackage{lipsum}

\usepackage{mdframed}
\usepackage{amsmath,amssymb}
\usepackage{lipsum}
\usepackage{bm}
\usepackage{graphicx}
\usepackage{graphics}
\usepackage{amsmath}
\usepackage{array}
\usepackage[caption=false]{subfig}
\usepackage{xcolor}
\usepackage{subfig}
\usepackage{hyperref}
\usepackage[capitalise]{cleveref}
\usepackage{textcomp}
\usepackage{epstopdf}
\usepackage{cases}
\usepackage{amssymb}
\usepackage{multirow}
\usepackage{siunitx}
\usepackage{cases}
\usepackage{tabularx}
\usepackage[utf8]{inputenc}
\usepackage{tikz}
\usepackage{comment}
\usepackage{lineno,hyperref}
\modulolinenumbers[5]
\usepackage{booktabs}

\journal{ELSEVIER}

\begin{document}

\begin{frontmatter}






\title{Effects of heating strategies and ballistic transport on the transient thermal conduction in 3D FinFETs}
\author[add1]{Chuang Zhang}
\ead{zhangc520@hdu.edu.cn}
\author[add2]{Ziyang Xin}
\ead{xinziyang@hust.edu.cn}
\author[add3]{Qin Lou}
\ead{louqin560916@163.com}
\author[add1]{Hong Liang\corref{cor1}}
\ead{lianghongstefanie@163.com}
\address[add1]{Department of Physics, School of Sciences, Hangzhou Dianzi University, Hangzhou 310018, China}
\address[add2]{School of Energy and Power Engineering, Huazhong University of Science and Technology, Wuhan, 430074, China}
\address[add3]{School of Energy and Power Engineering, University of Shanghai for Science and Technology, Shanghai 200093, China}
\cortext[cor1]{Corresponding author}
\date{\today}

\begin{abstract}
Efficiently capturing the three-dimensional spatiotemporal distributions of temperature is of great significance for alleviating hotspot issues in 3D FinFETs. 
However, most previous thermal simulations mainly focused on the steady-state problem with continuous heating.
Few studies are conducted for the transient thermal conduction in 3D FinFETs with non-continuous heating, which is actually closer to the reality.
To investigate the effects of heating strategies on the transient micro/nano scale thermal conduction in 3D FinFETs, three heating strategies are considered, including `Continuous', `Intermittent' and `Alternating' heating.
A comparison is made between the phonon BTE solutions and the data predicted by the macroscopic diffusion equation, where the effect of boundary scattering on phonon transport is added into the effective thermal conductivity.
Numerical results show that it is not easy to accurately capture the heat conduction in 3D FinFETs by the macroscopic diffusion equation, especially near the hotspot areas where ballistic phonon transport dominates and the temperature diffusion is no longer valid.
Different heating strategies have great influence on the peak temperature rise and transient thermal dissipation process.
Compared to `Intermittent' or `Continuous' heating, the temperature variance of `Alternating' heating is smaller.
\end{abstract}




\end{frontmatter}




\section{INTRODUCTION}

Advanced process technologies represented by Fin field-effect transistors (FinFETs) continue to promote the development of the semiconductor and microelectronics industries~\cite{IRDS2023,Fin_GAAFET_2024_review,heat_dissipations_cross_Scale_review_2019,pop2004transistors}. 
However, when the number density, heat generation power and heat flux density of transistors increases sharply, the hotspot problem becomes increasingly serious, threatening the safe operation of electronic devices~\cite{experiments_2020_transient,English_rancheng_low_temperature_FinFET2024,WANG2024121499}. 
As a mainstream structure in logic chips or semiconductor devices, single three-dimensional FinFET unit has geometric dimensions ranging from tens of nanometers to hundreds of nanometers.
This size scale is comparable to the phonon mean free path of room temperature silicon, so that the classic Fourier law of heat conduction no longer holds~\cite{KIM2019114080,ziabari2018a,RevModPhys.90.041002,ZHANG2020}.
Therefore, it is very important to understand the micro/nano scale heat transfer physics and efficiently predict the three-dimensional temporal and spatial distributions of temperature in order to alleviate the hotspot problem in microelectronics~\cite{TCAD_application_intel_2021_review,pop_energy_2010,2024_VLSI_thermal_IMEC,BEOL2023_IMEC,ZHANG2024123379}.

A mainstream engineering treatment method is to adopt a 3D heat diffusion equation with an effective thermal conductivity which takes into account the size effect~\cite{2024_VLSI_thermal_IMEC,BEOL2023_IMEC,IEEE_qinghao_2018,TCAD_application_intel_2021_review,intel_2023_GAAFET,ZHANG2025126374,shen2024_APL}.
This method is widely used in many industrial softwares such as TCAD, ANSYS and COMOSL.
These softwares contain huge material databases, so that the users can select the appropriate thermal conductivity coefficient based on the material components, size, temperature, doping concentration, etc. 
This also indicates that its numerical accuracy significantly depends on the engineers' experience.
The disadvantage is that it still assumes a linear relationship between the heat flux and temperature gradient.
To solve this drawback, many macroscopic moment equations~\cite{PhysRevB.103.L140301,RevModPhysJoseph89,KOVACS20241,hydrodynamicsHoussem2021,ziabari2018a} have been developed by introducing time delay terms, memory effects, heat flux nonlinearity or nonlocal terms, that is, introducing high-order derivatives or tensors of temperature and heat flux with respect to time and space.
These non-Fourier heat conduction models can capture non-equilibrium or non-diffusive effects to some extent.
However, a small parameter expansion assumption is usually adopted during derivation process, so that it is difficult for most of them to accurately characterize ballistic transport, especially in complex 3D geometries.

Another accessible method is to numerically solve the mesoscopic phonon Boltzmann transport equation (BTE)~\cite{ZHANG2023124715,TCAD_application_intel_2021_review,mazumder_boltzmann_2022,barry2022boltzmann}.
Adisusilo, $et~al$ used the Monte Carlo method to simulate the thermal properties of a 3D bulk FinFET with gate length $22$ nm and fin thickness $8$ nm.
Numerical results show that the peak temperature predicted by classical Fourier's law is $100$ K lower than that of Monte Carlo method.
Hu, $et~al$~\cite{HU2024huabao} and Zhang, $et~al$~\cite{ZHANG2025126374} used the synthetic iterative scheme to capture non-equilibrium thermal conduction in 3D bulk FinFET.
Numerical results show that the peak temperature rise in FinFET predicted by the phonon BTE deviates from those predicted by the macroscopic diffusion equation even if an effective thermal conductivity is used~\cite{ZHANG2025126374}.

The phonon BTE is usually regarded as the core bridge in connecting microscopic or macroscopic methods in many multi-scale thermal simulation~\cite{coupled_samsung2018}.
For instance, the IMEC research team~\cite{2023IEEEthermal_IMEC,BEOL2023_IMEC,2024_VLSI_thermal_IMEC} used the modular method to evaluate the thermal performance of the back-end of line.
They calculated the thermal physical parameters of electrons and phonons using the first principles as the input parameters of BTE, and then used the Monte Carlo method to numerically solve the BTE and extracted the effective thermal conductivity of materials such as nanoscale interconnects and through-holes. 
Finally, the macroscopic diffusion equation is solved to extract the effective thermal resistance of each layer of back-end of line in a reasonable calculation time. 
The Intel research team~\cite{TCAD_application_intel_2021_review,intel_2023_GAAFET} numerically solves the phonon BTE to obtain the effective thermal conductivity of the entire silicon fin or nanowires, where the heat source term is obtained from electron-phonon coupling.
After that, a larger thermal simulations at the cell or circuit level is conducted to assess the effects of self-heating on interconnects and circuits by numerically solving the macroscopic equation.

In the above studies of heat dissipation in FinFETs~\cite{PhysRevApplied.19.014007,baohua_IEEE_2024,intel_2023_GAAFET,BEOL2023_IMEC,3DFINFET_2014_mc}, a continuous heating source is usually used and steady-state temperature field is analyzed.
However, the electronic equipments do not always work and the transient thermal evolution process is much more noteworthy in reality~\cite{TSMC_2018_self_heating,TCAD_application_intel_2021_review,PhysRevE.106.014111}.
In this paper, the effects of heating strategies on the transient thermal conduction in FinFET are investigated based on the phonon BTE.
Three heating strategies are accounted and the associated steady or unsteady thermal dissipations processes are simulated, analyzed and discussed.
The rest of this article is organized as follows.
Theoretical models and methods are introduced briefly in Sec.~\ref{sec:BTE}. 
Results and discussions for the heat dissipation in FinFETs are conducted in Sec.~\ref{sec:results}. 
Finally, conclusion and outlook are made in Sec.~\ref{sec:conclusion}.

\section{MODELS}
\label{sec:BTE}

The phonon BTE under the single-mode relaxation time approximation (RTA) is used to describe the phonon transport in FinFETs~\cite{GuoZl16DUGKS,ZHANG2023124715,TANG2023123497,TANG2023123497,MurthyJY05Review,mazumder_boltzmann_2022,barry2022boltzmann,intel_2023_GAAFET,warzoha_applications_2021},
\begin{align}
\frac{\partial f}{\partial t }+ v_g \bm{s} \cdot \nabla_{\bm{x}} f  = \frac{f^{eq}  -f}{\tau  } + \frac{1}{4 \pi} \dot{S} , \label{eq:BTE} 
\end{align}
where $f=f(\bm{x}, \bm{s},t )$ is the phonon distribution function of energy density, which depends on spatial position $\bm{x}$, unit directional vector $\bm{s}$ and time $t$.
$f^{eq} $ is the associated equilibrium distribution function, $v_g$ is the phonon group velocity, $\tau$ is the relaxation time and $\dot{S}=\dot{S}(\bm{x},t)$ is the external heat source.
Taking an integral of Eq.~\eqref{eq:BTE} over the solid angle space $d\Omega$ leading to the first law of thermodynamics with energy $U$ and heat flux $\bm{q}$,
\begin{align}
\frac{\partial U}{\partial t }+ \nabla_{\bm{x}} \cdot \bm{q} =  \int  \frac{f^{eq} -f}{\tau }   d\Omega  + \dot{S} = \dot{S} , \label{eq:firstlaw} \\
U =  \int f d\Omega = C T,  \quad  \quad  \bm{q} =  \int v_g \bm{s} f d\Omega. 
\end{align}
where $C$ is the specific heat and $T$ is the temperature.
The phonon scattering kernel satisfies energy conservation,
\begin{align}
\int \frac{f^{eq}  -f}{\tau  } d\Omega =0.
\end{align}
Silicon and silicon dioxide are two main semiconductor materials in FinFETs, and their thermal properties at room temperature $300$ K can be obtained from previous references~\cite{Silicon-on-Insulator_1992_IEEE,YangRg05BDE,NASRI2015206}, as listed in Table.~\ref{Si_SiO2_parameters}.

\renewcommand\arraystretch{1.2}
\begin{table}[htb]
\caption{Thermal properties of room temperature silicon (Si) and silicon dioxide (SiO$_2$)~\cite{Silicon-on-Insulator_1992_IEEE,YangRg05BDE,NASRI2015206}.}
\centering
\begin{tabular}{*{4}{c}}
\hline
\hline
  & $C$ (J$\cdot$ m$^{-3} \cdot $K$^{-1}$)  &  $v_g$ (m$\cdot$s$^{-1}$)  &  $\lambda$ (nm)    \\
\hline
Si   & 1.5E6 & 3.0E3  &  100.0       \\
\hline
SiO$_2$   & 1.75E6 & 5.9E3  &  0.4      \\
\hline
\hline
\end{tabular}
\label{Si_SiO2_parameters}
\end{table}

Discrete unified gas kinetic scheme (DUGKS)~\cite{GuoZl16DUGKS,zhang_discrete_2019} is used to solve the phonon BTE accounted for the interfacial thermal resistance~\cite{dengke_2024_APL,RevModPhys.94.025002,JAP_qinghao_2017,IEEE_qinghao_2018}.
Detailed introductions and numerical settings can be found in~\ref{sec:dugks} and~\ref{sec:discretization}.
Note that the numerical discrete solution of the seven-dimensional phonon BTE requires extremely high computing resources and cost. Hence, the isotropic wave vector and frequency-independence assumption is used in the present paper to achieve a good balance between computational efficiency and accuracy.
All numerical simulations are conducted by a home-made C/C++ program~\cite{zhang_discrete_2019,PhysRevE.106.014111}.

\section{RESULTS AND DISCUSSIONS}
\label{sec:results}

\begin{figure}[htb]
\centering  
\includegraphics[width=0.48\textwidth]{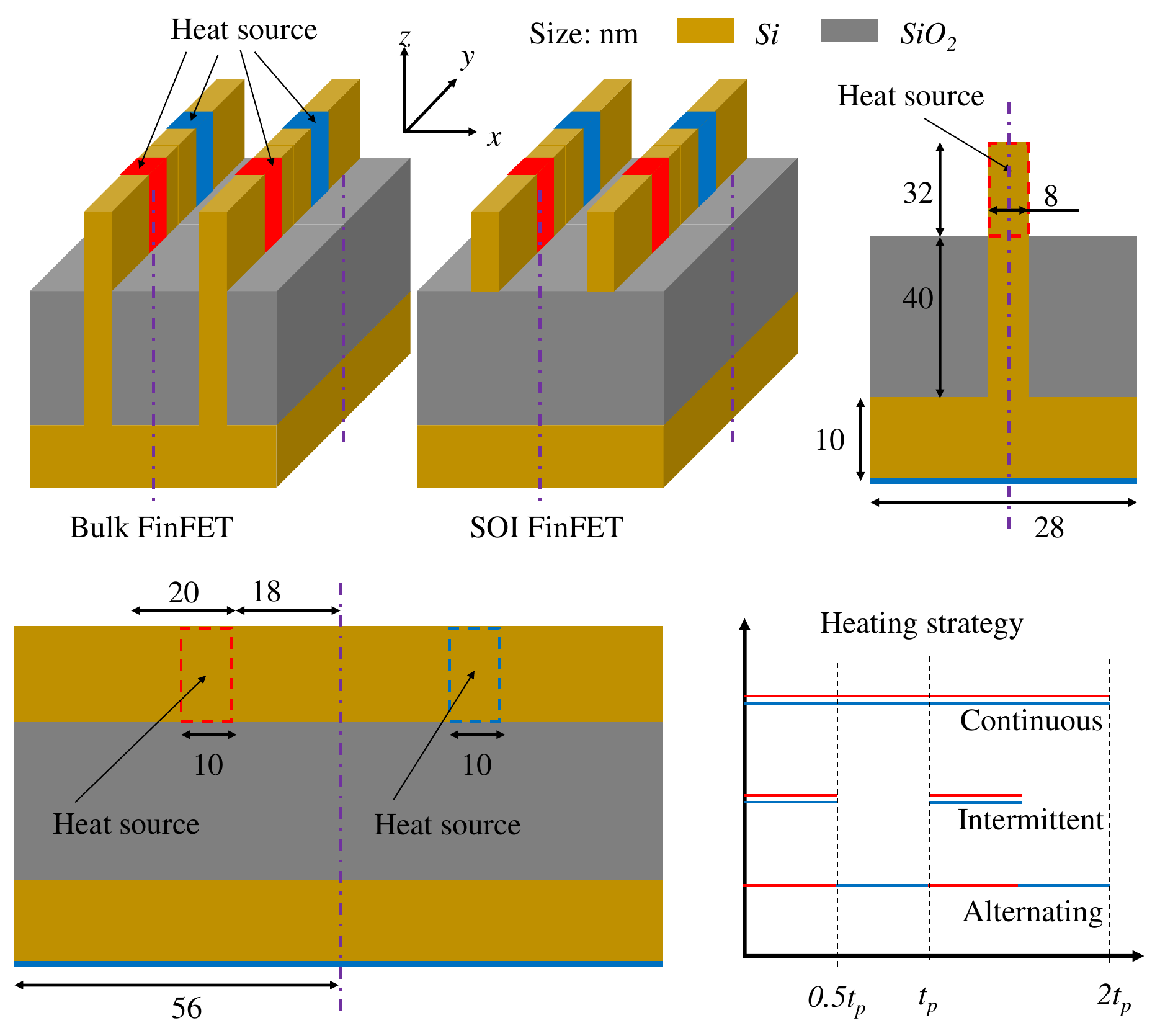}
\caption{Three-dimensional global graph of four bulk or silicon-on-insulator (SOI) FinFETs. For the red and blue heat source areas in 3D FinFETs, the following three heating strategies are used, including `Continuous', `Intermittent', `Alternating', where $t_p$ is a heating period. }
\label{quasi3dhot}
\end{figure}

Motivated by previous thermal simulations of 3D FinFET~\cite{3DFINFET_2014_mc,3DFINFETtransient,intel_2023_GAAFET,baohua_IEEE_2024,ZHANG2025126374} based on the phonon BTE, we only investigates the three-dimensional transient phonon conduction problem in single or several FinFETs in this paper.
Schematics of 3D bulk or SOI FinFETs are shown in~\cref{quasi3dhot}, both of which are composed of silicon fin, silicon dioxide insulation layer and silicon substrate with several contact interfaces.
The system sizes in the $x$, $y$, $z$ direction of a single FinFET are $28$nm, $56$ nm and $82$ nm, respectively.
Front, back, left and right surfaces are all symmetric boundaries. 
The whole diagram is geometrically symmetric in the $x$ and $y$ directions with respect to the purple dot dash lines.
Bottom surfaces are the heat sink with fixed environment temperature and the other surfaces are all diffusely reflecting adiabatic boundaries.
The heat source is located in the fin area, whose system size in the $x$, $y$, $z$ direction are $8$, $10$ and $32$ nm, respectively.
The entire structure size is approximately a $7$ nm technology node transistor.

Three heating strategies are mainly considered, where `Continuous' represents that the two external heat sources always heat the system. 
It indicates a steady heat source, which is used in most previous papers.
`Intermittent' represents that the two external heat sources work together and both of them heat the system in a half time. 
It is a bit like the heating method in pump-probe experiments, where the heat source heats the system for a while and does not heat it for a while.
`Alternating' represents that the two external heat sources work alternatively and each one heats the system in a half time, where $t_p$ is a heating period.
It is much like the N and P type transistors in chips are periodically arranged on the substrate and interlace to work when the AC voltage is loaded.
For typical chips in electronic devices (e.g., laptop), the working frequency is about $4$ GHz so that we set $t_p=0.25$ ns.
The maximum heating power is $5.0 \times 10^{16}$ W$\cdot$m$^{-3}$.
The temperature dependence of phonon thermophysical parameters~\cite{zhang_discrete_2019,ZHANG2025126374} is not accounted, therefore, the predicted temperature rise is basically linear to the input power density based on dimensionless analysis of the phonon BTE~\cite{GuoZl16DUGKS}.

A lot of previous experimental or theoretical work has proven the failure of classical Fourier's law at the micro/nano scale~\cite{RevModPhys.90.041002,honarvar_directional_2021,PhysRevApplied.10.054068,ZHANG2020}, so we focus on comparing the BTE results with the macroscopic diffusion equation with empirically modified effective thermal conductivity coefficients~\cite{2024_VLSI_thermal_IMEC,BEOL2023_IMEC,IEEE_qinghao_2018,TCAD_application_intel_2021_review,intel_2023_GAAFET,ZHANG2025126374,shen2024_APL}, namely, effective Fourier's law (EFL).
Detailed introductions and numerical discretizations can be found in~\ref{sec:diffusionsolver}.

\begin{figure}[htb]
\centering  
\includegraphics[width=0.45\textwidth]{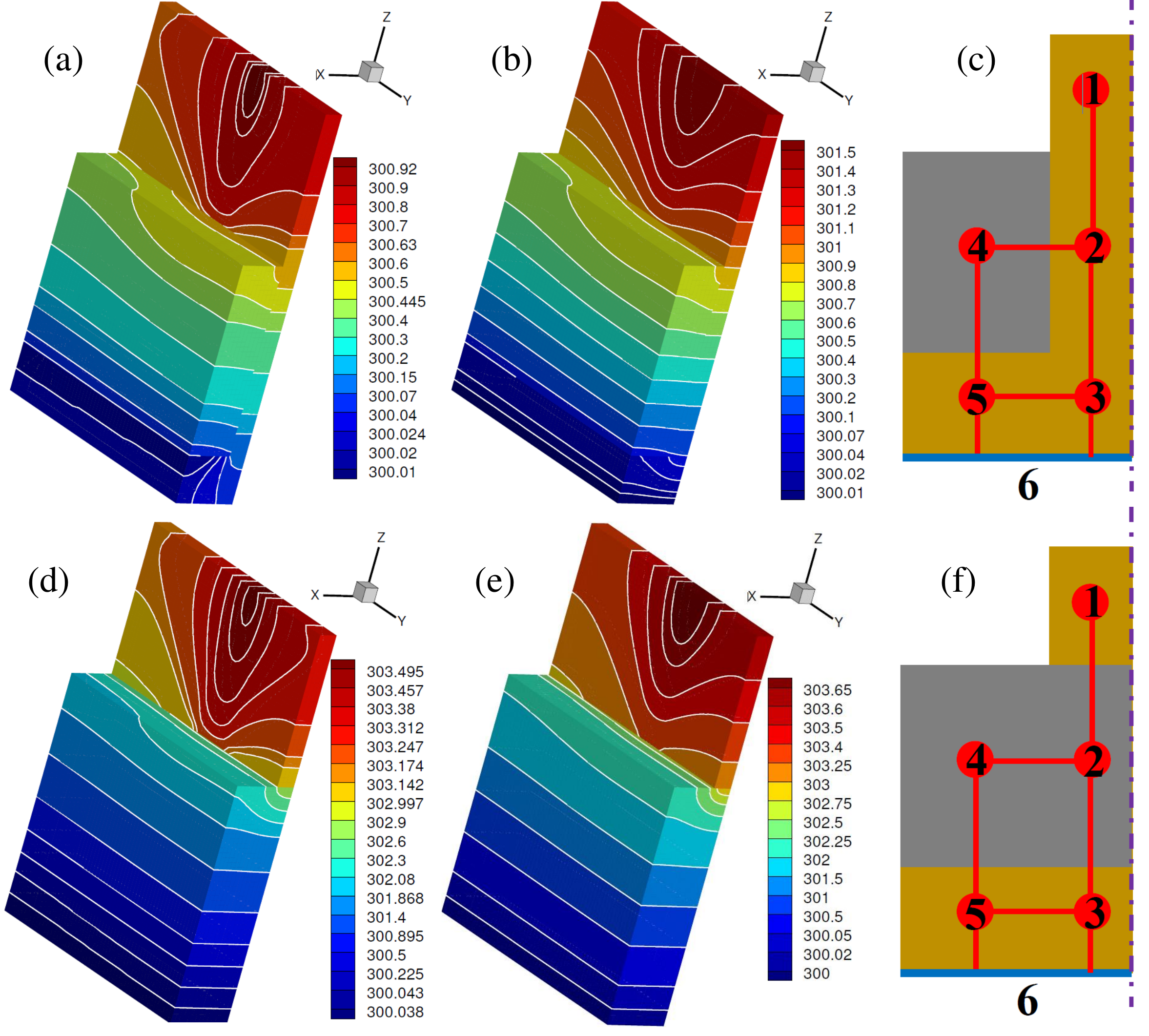}
\caption{The first two column are the steady temperature contour of a half of FinFET under `Continuous' heating. The last column is the schematic of heat dissipation path from the heat source to heat sink in the $XZ$ plane. (a,b,c) Bulk FinFET. (d,e,f) SOI FinFET. (a,d) BTE. (b,e) EFL. }
\label{3DFinFET_steady}
\end{figure}
\begin{figure}[htb]
\centering  
\subfloat[Bulk FinFET]{\includegraphics[width=0.4\textwidth]{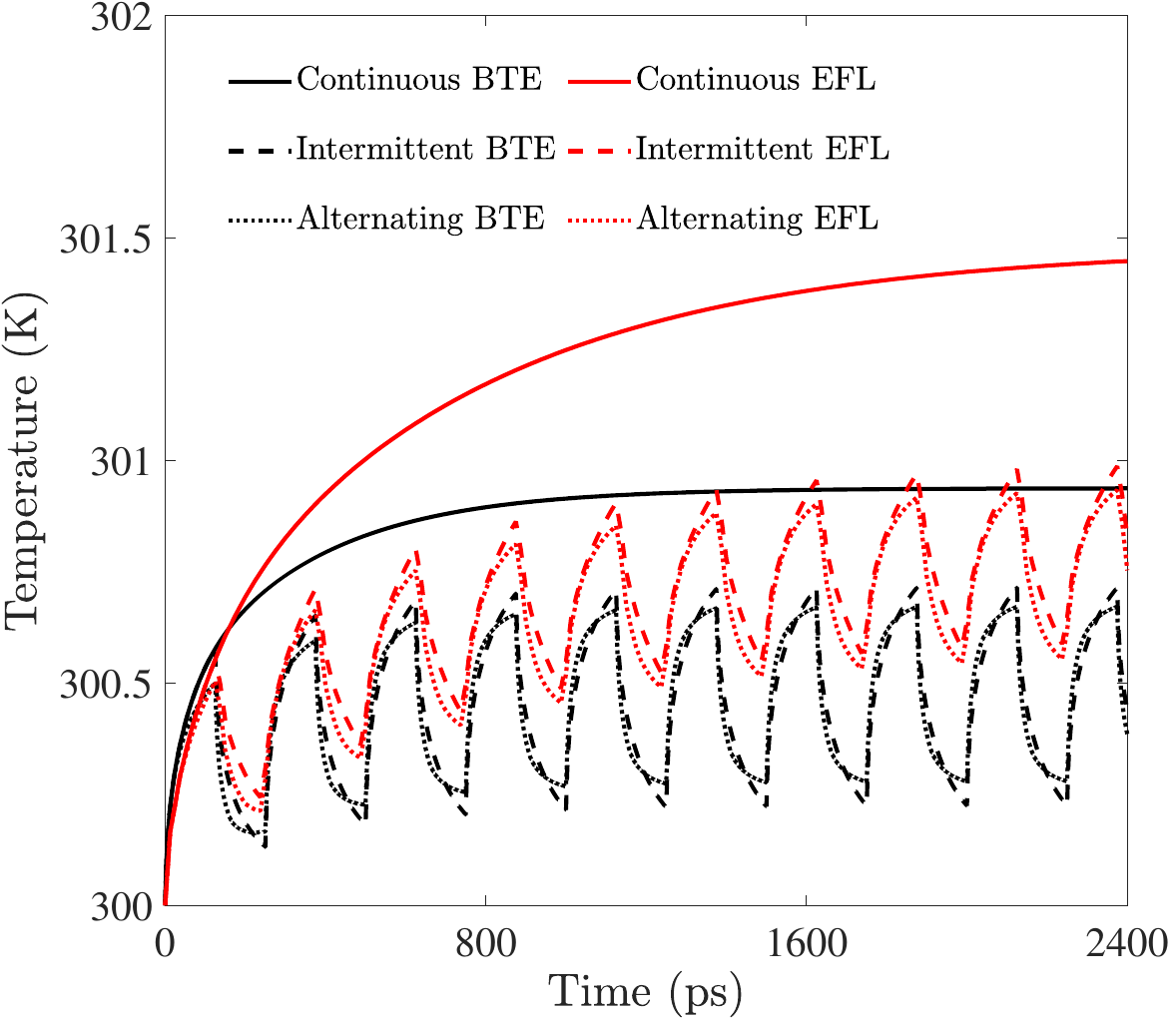}} \\
\subfloat[SOI FinFET]{\includegraphics[width=0.4\textwidth]{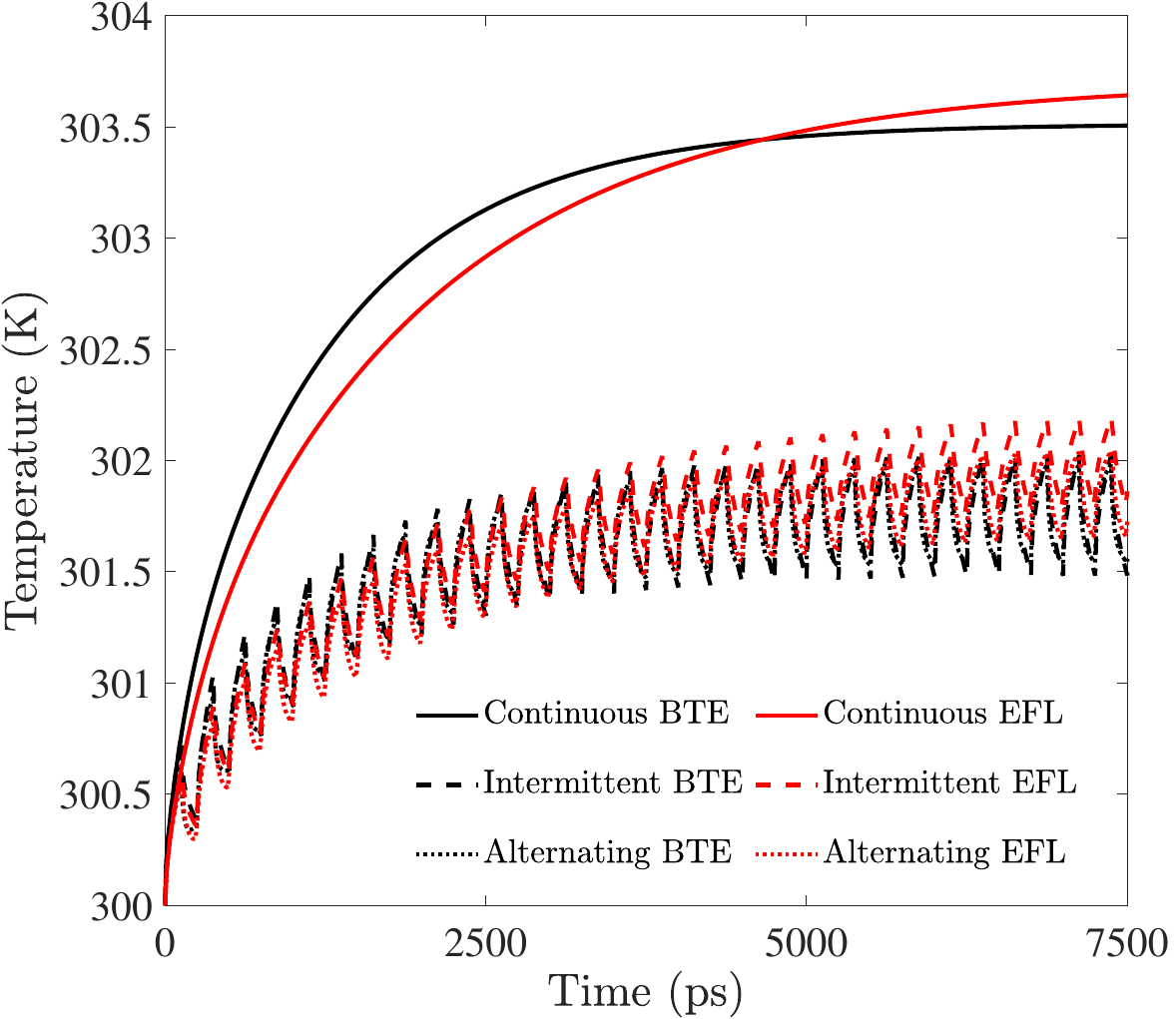}} \\
\caption{Transient evolutions of the maximum temperature in the 3D hotspot system under three heating strategies.}
\label{3DFinFET_transient}
\end{figure}
\begin{figure}[htb]
\centering  
\subfloat[$t=0.05t_p$]{\includegraphics[scale=0.20,viewport=90 10 450 530,clip=true]{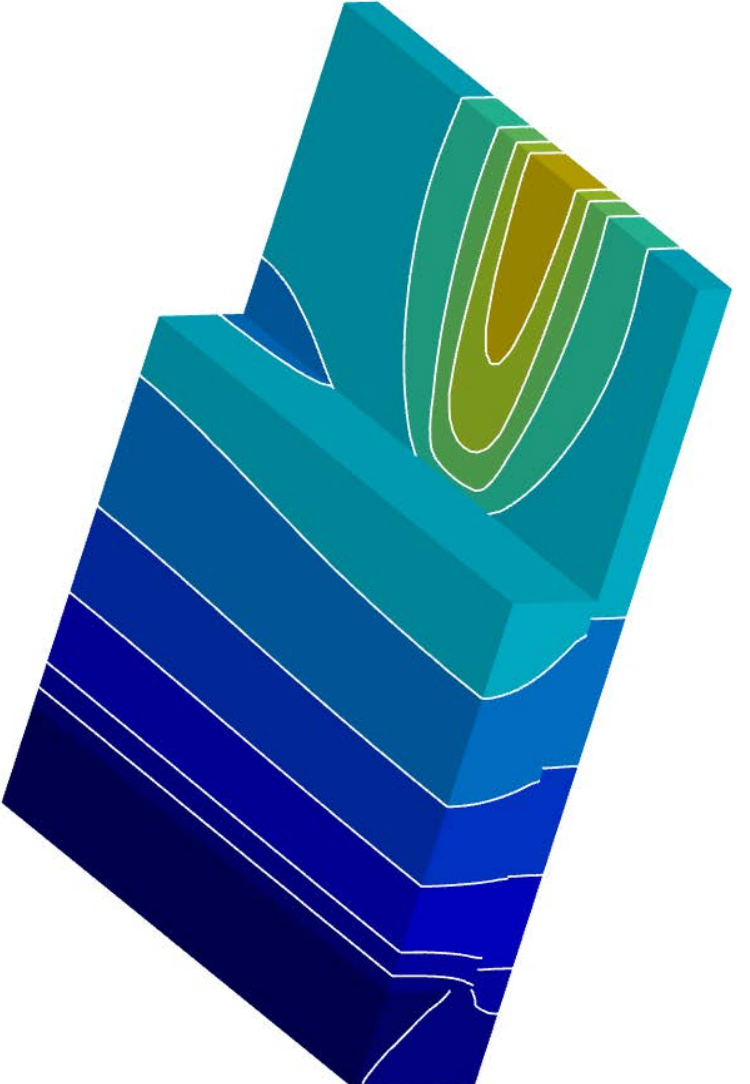}} 
\subfloat[$t=0.25t_p$]{\includegraphics[scale=0.20,viewport=90 10 450 530,clip=true]{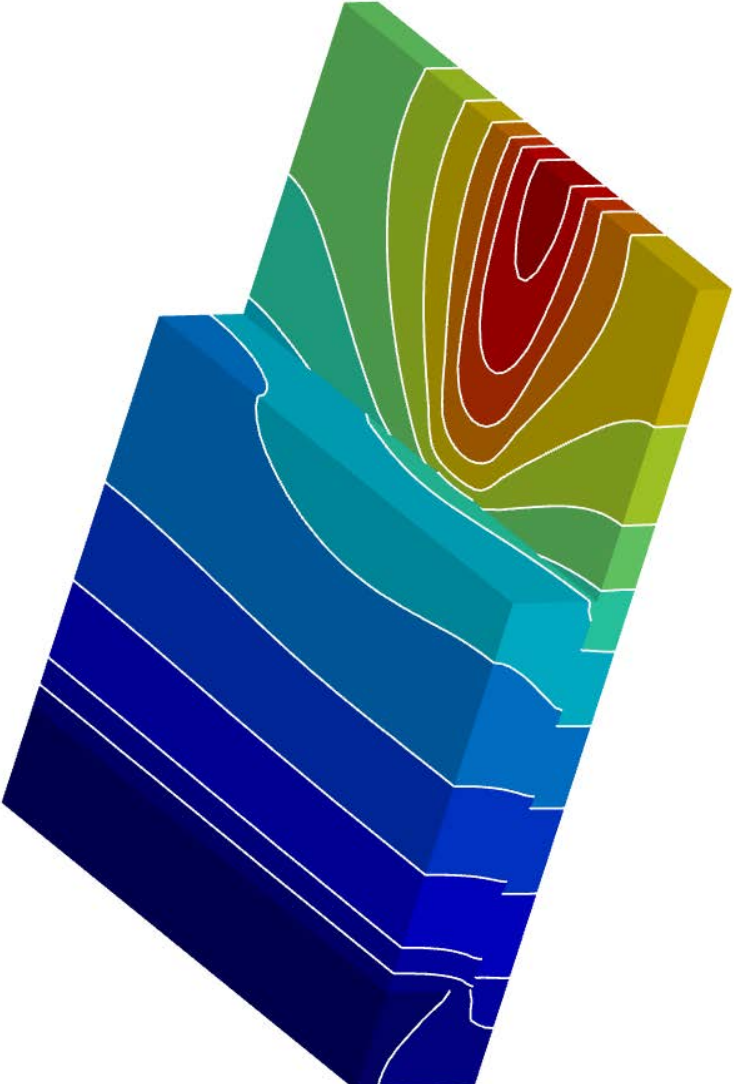}} 
\subfloat[$t=0.5t_p$]{\includegraphics[scale=0.20,viewport=90 10 450 530,clip=true]{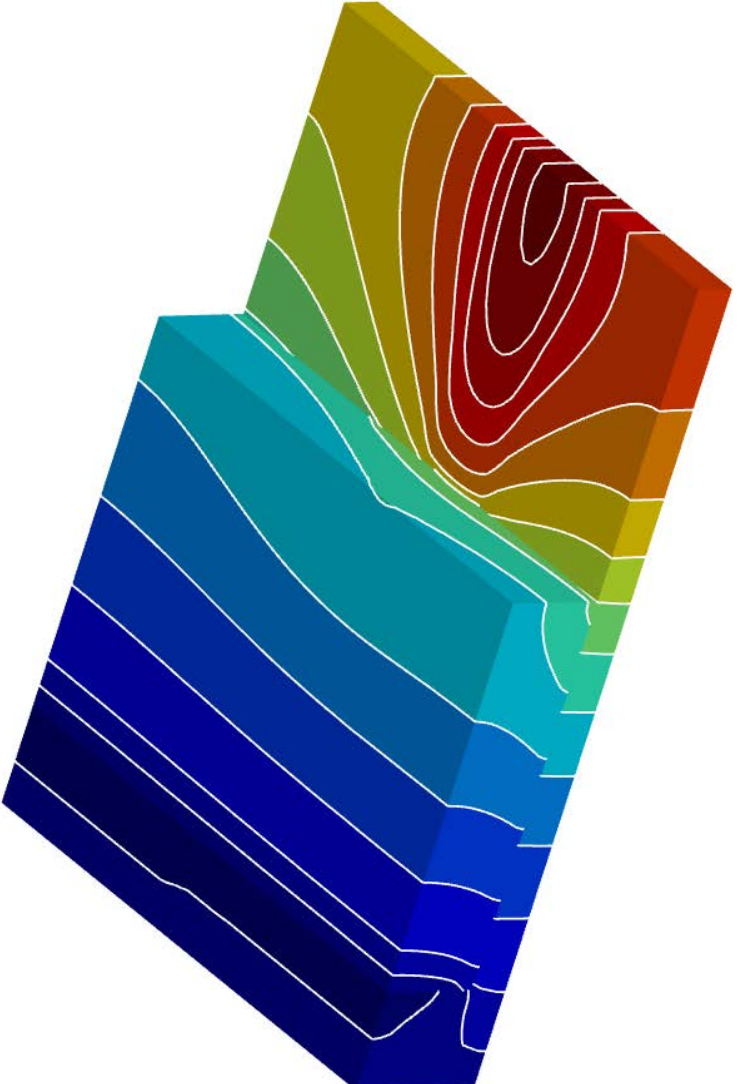}} \\
\subfloat[$t=0.55t_p$]{\includegraphics[scale=0.20,viewport=90 10 450 530,clip=true]{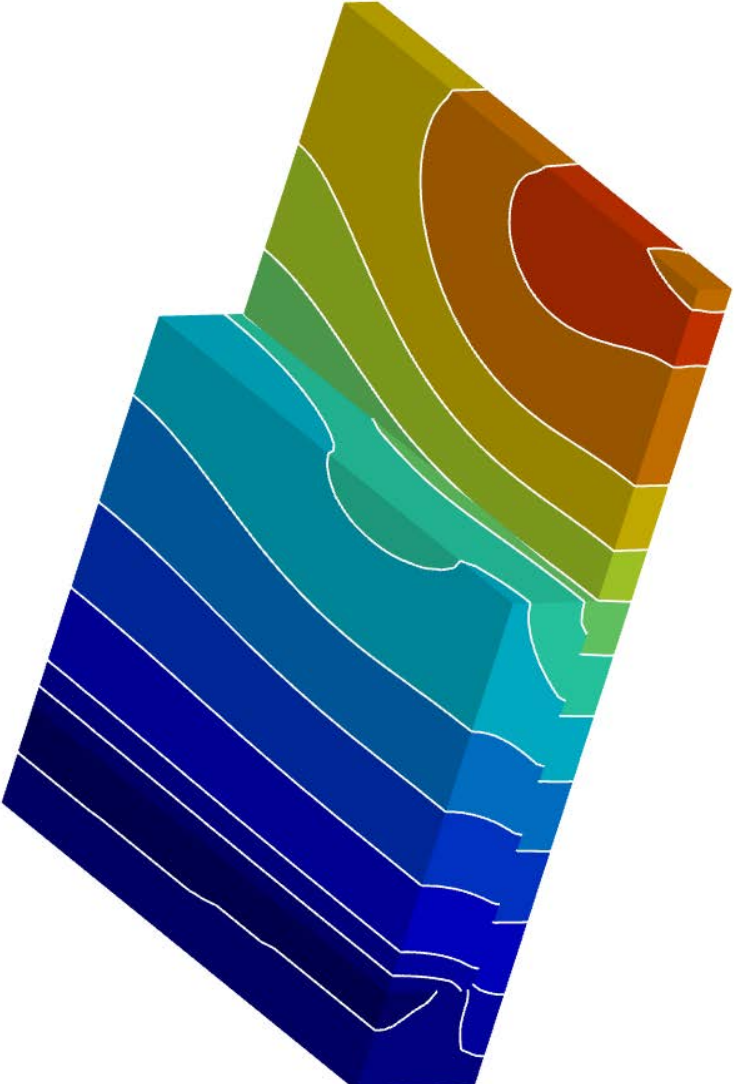}}
\subfloat[$t=0.75t_p$]{\includegraphics[scale=0.20,viewport=90 10 450 530,clip=true]{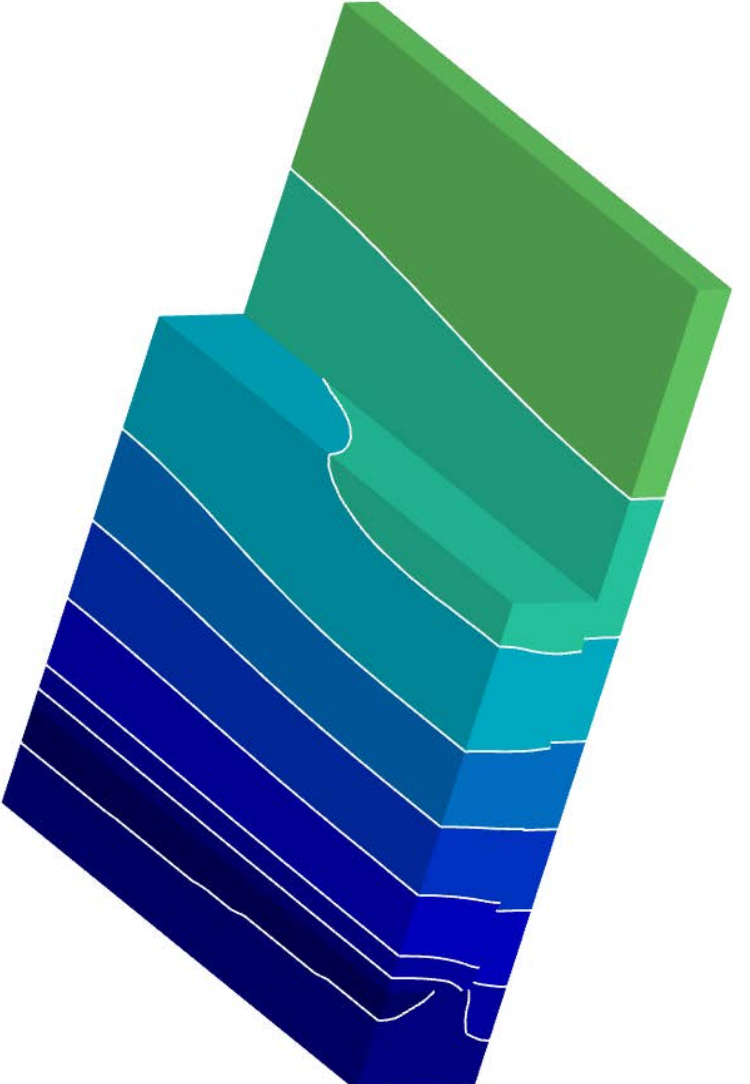}} 
\subfloat[$t=1.00t_p$]{\includegraphics[scale=0.20,viewport=90 10 530 530,clip=true]{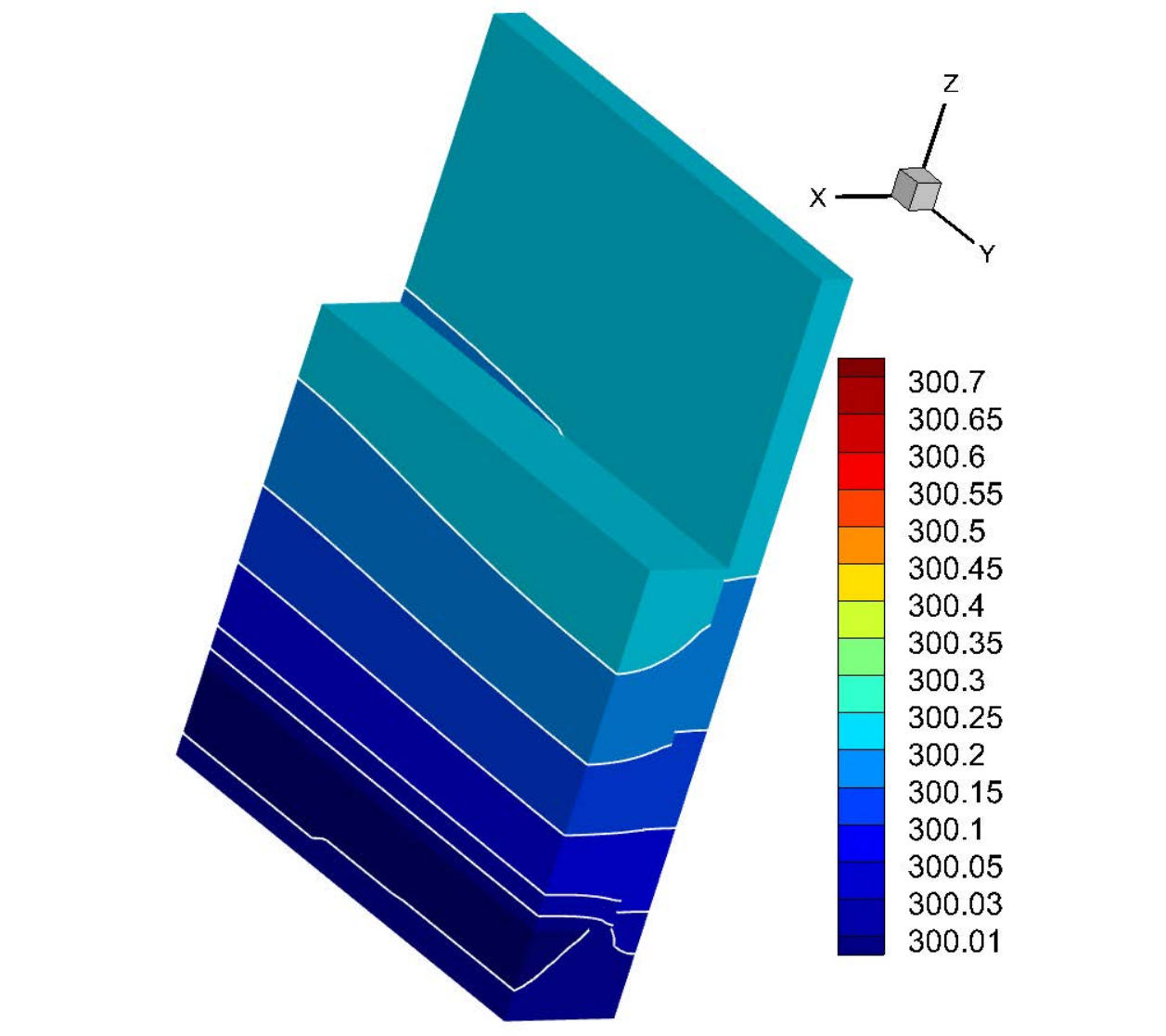}} 
\caption{Transient temperature contour of a half of bulk FinFET predicted by BTE under `Intermittent' heating when the system reaches the periodic steady state. }
\label{bulkFinFET_transient}
\end{figure}
\begin{figure}[htb]
\centering  
\subfloat[$t=0.05t_p$]{\includegraphics[scale=0.19,viewport=80 10 480 530,clip=true]{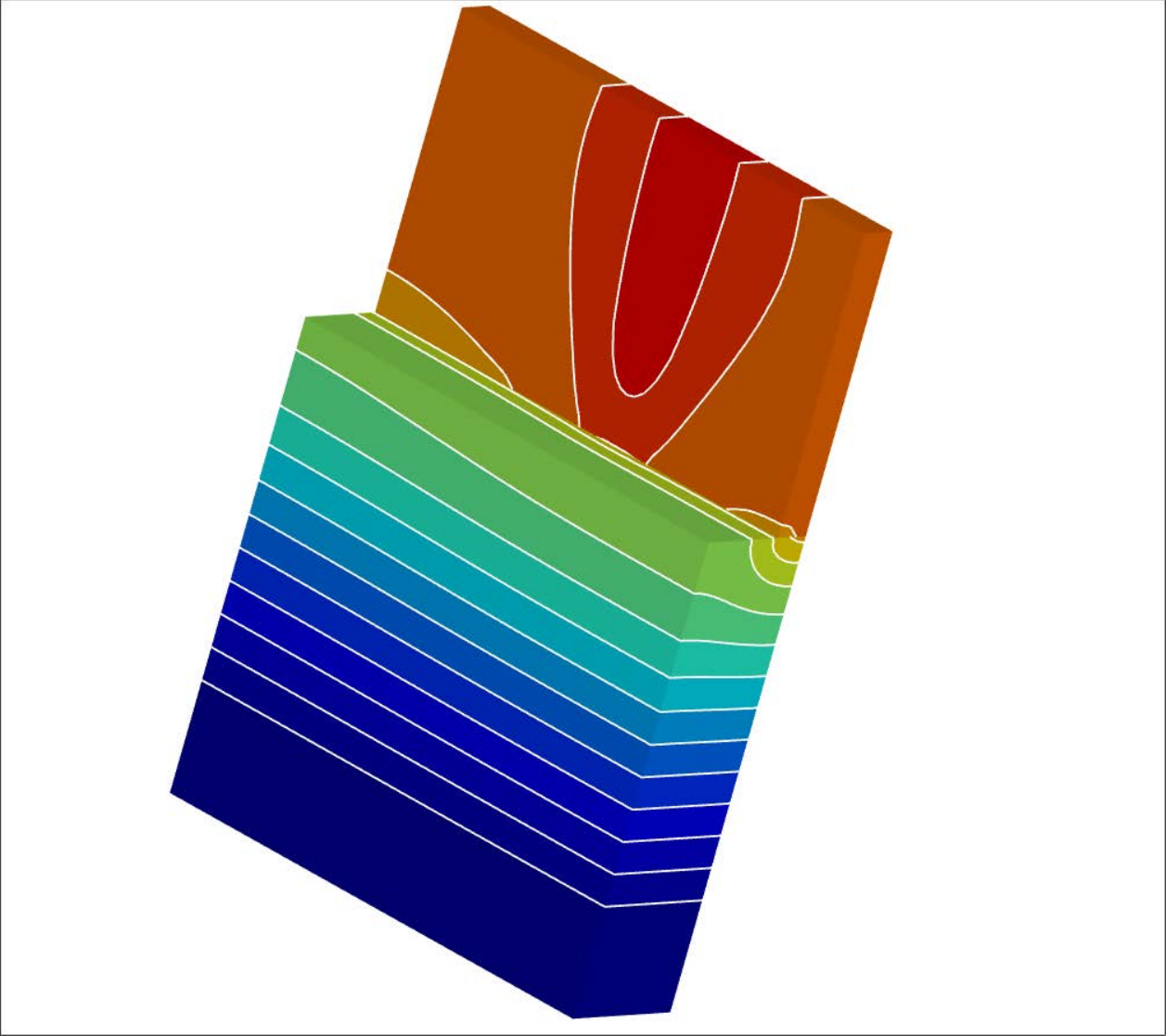}} 
\subfloat[$t=0.25t_p$]{\includegraphics[scale=0.19,viewport=80 10 480 530,clip=true]{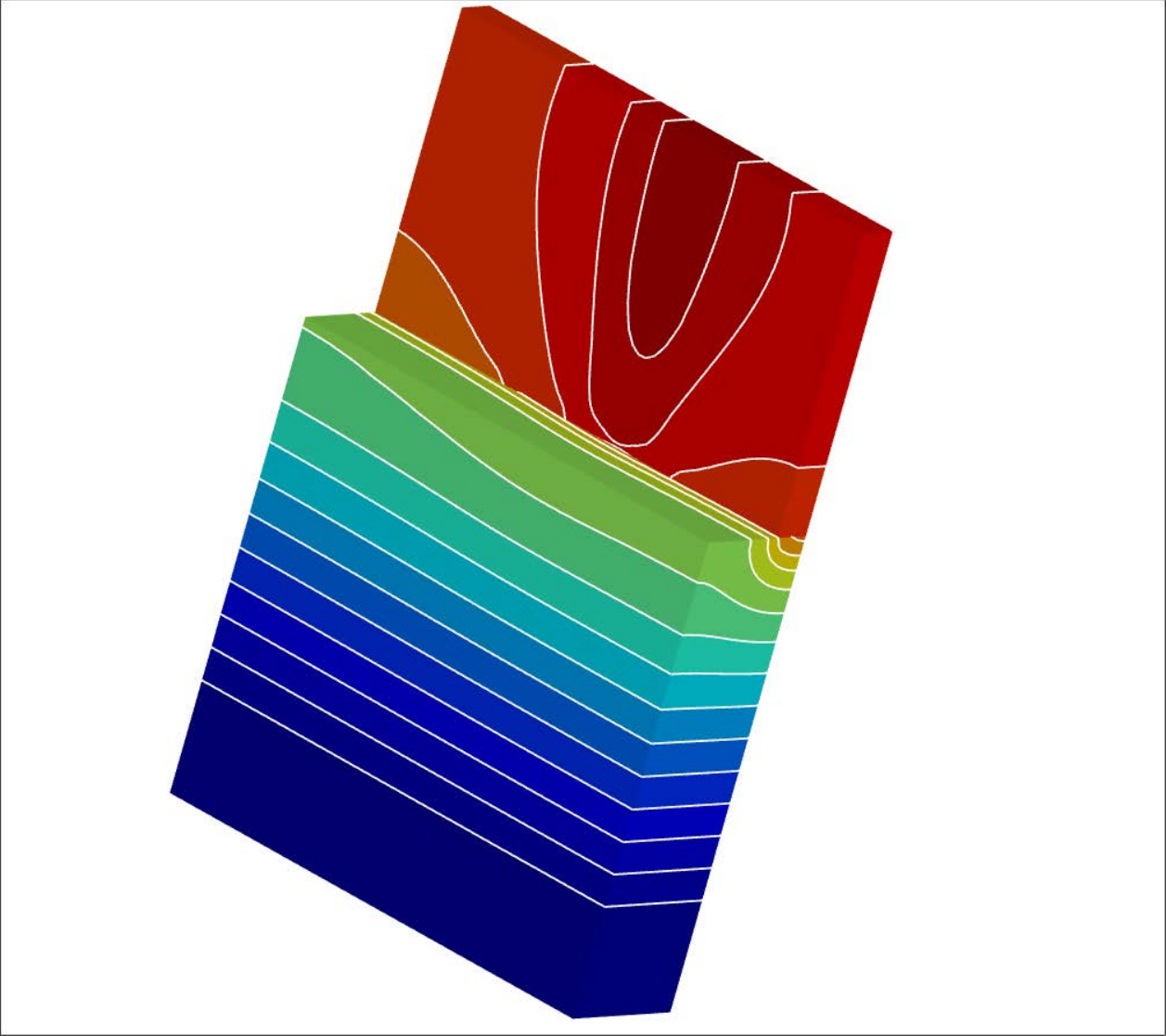}} 
\subfloat[$t=0.5t_p$]{\includegraphics[scale=0.19,viewport=80 10 480 530,clip=true]{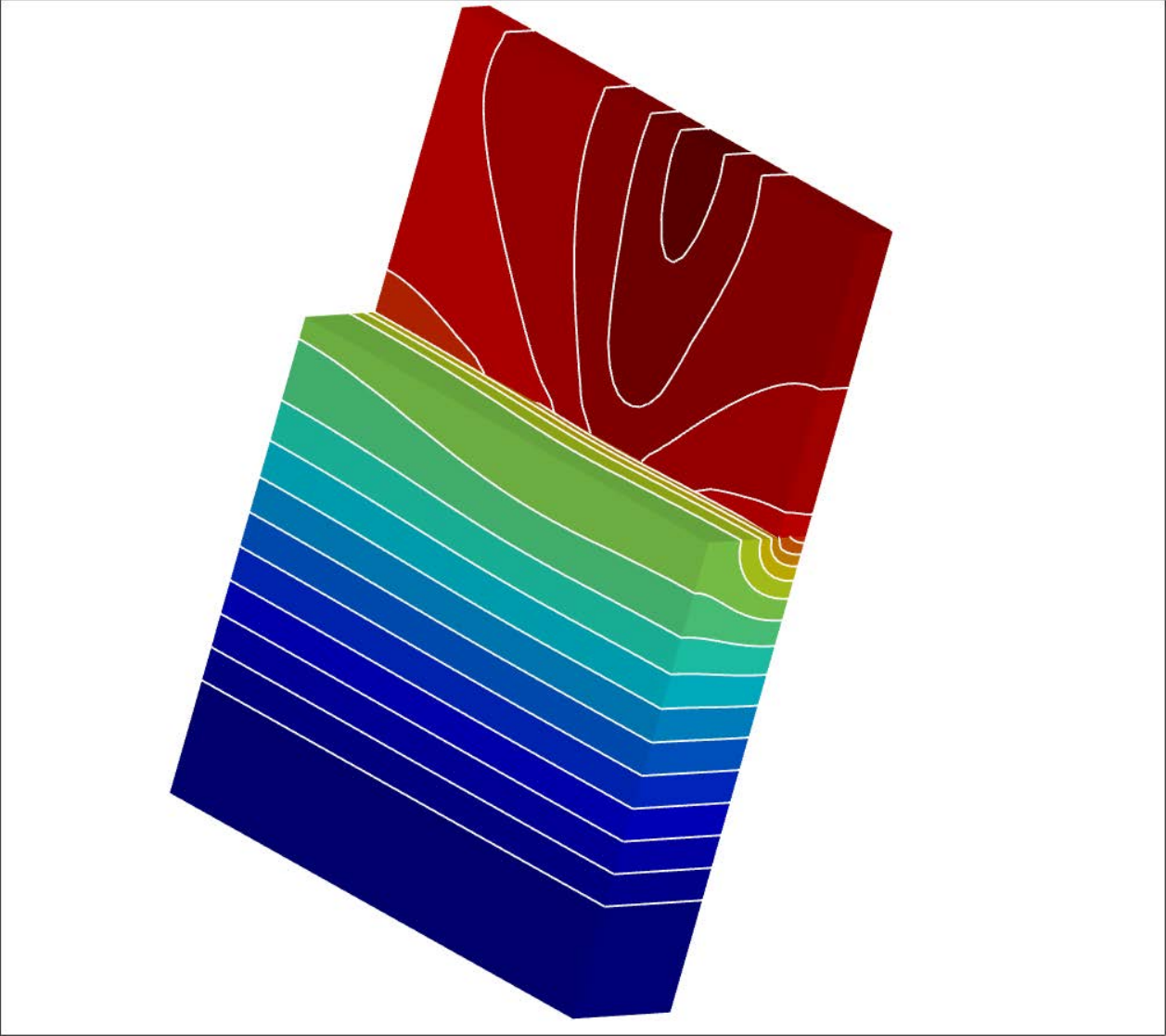}} \\
\subfloat[$t=0.55t_p$]{\includegraphics[scale=0.19,viewport=80 10 480 530,clip=true]{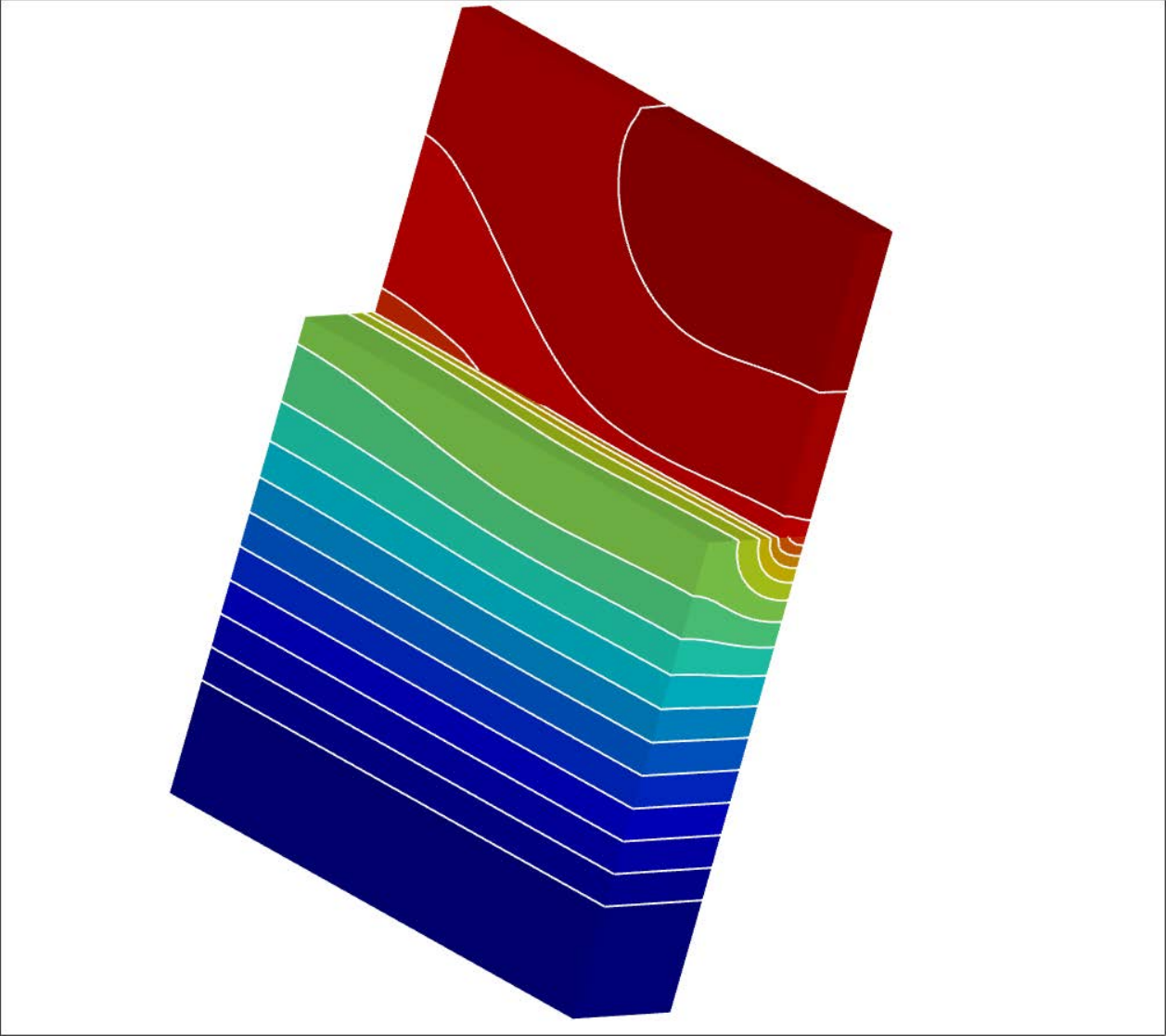}} 
\subfloat[$t=0.75t_p$]{\includegraphics[scale=0.19,viewport=80 10 480 530,clip=true]{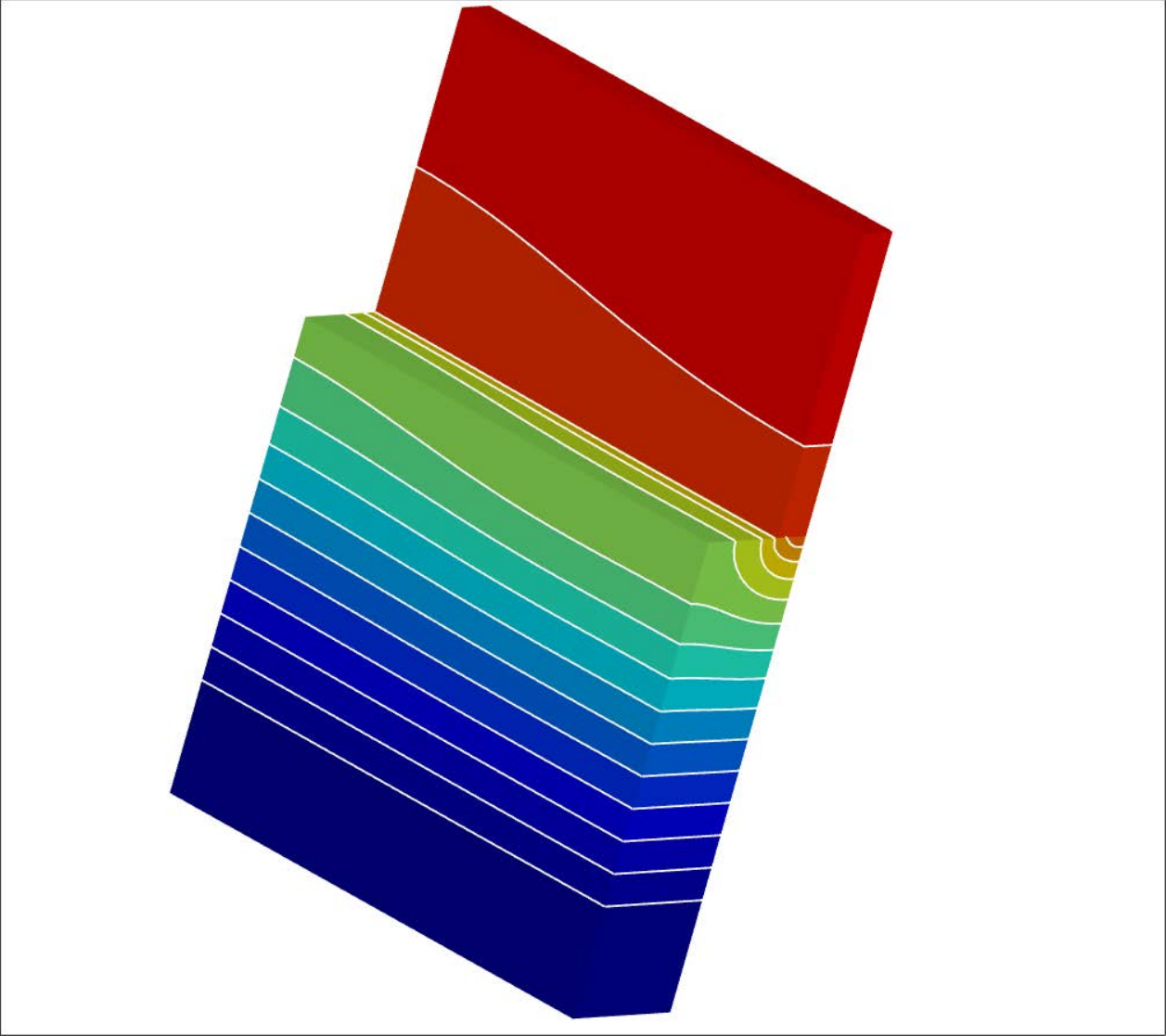}} 
\subfloat[$t=1.00t_p$]{\includegraphics[scale=0.19,viewport=80 10 550 530,clip=true]{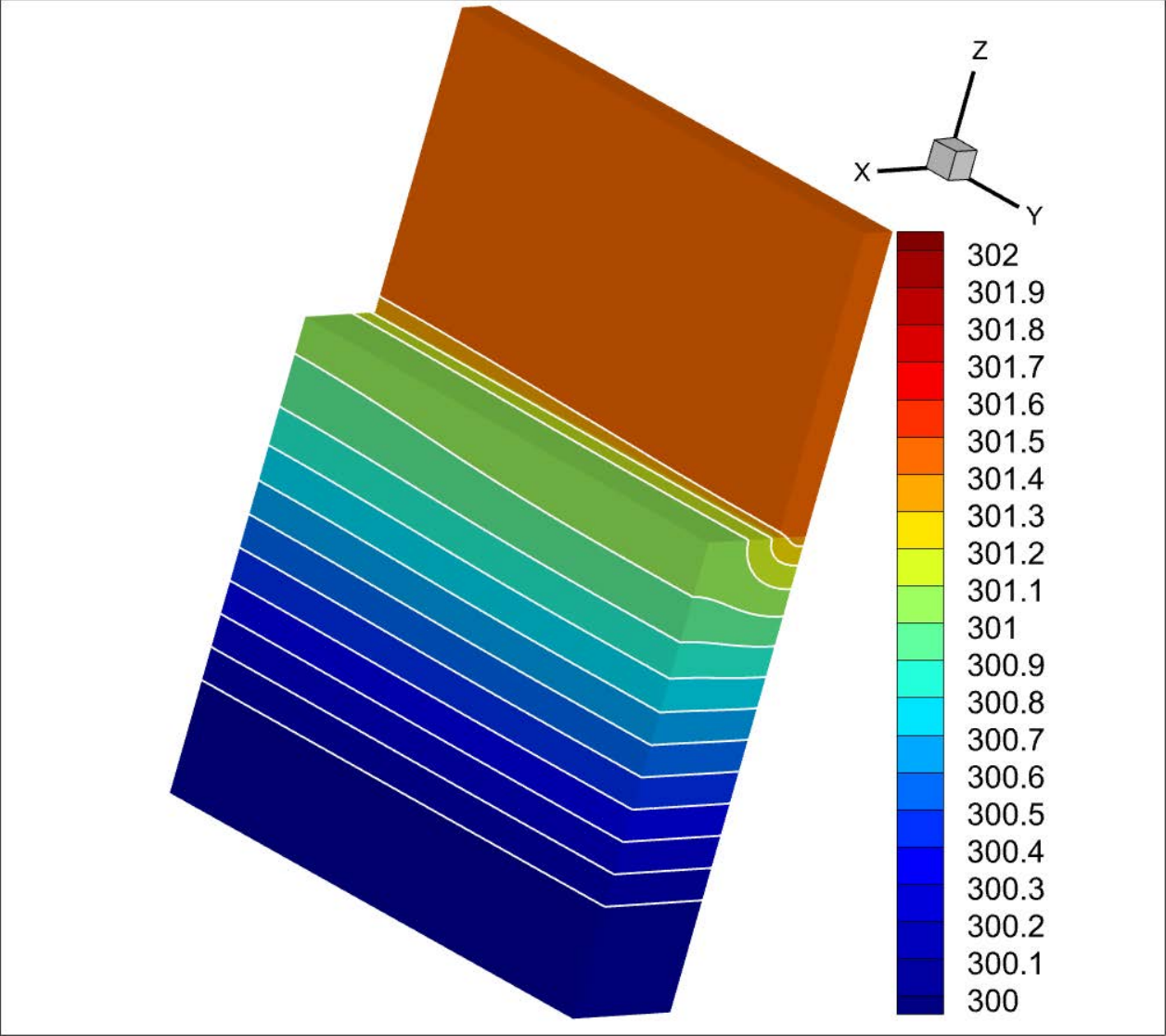}} 
\caption{Transient temperature contour of a half of SOI FinFET predicted by BTE under `Intermittent' heating when the system reaches the periodic steady state. }
\label{SOIFinFET_transient}
\end{figure}
\begin{figure}[htb]
\centering  
\subfloat[$t=0.05t_p$]{\includegraphics[scale=0.17,viewport=50 10 500 530,clip=true]{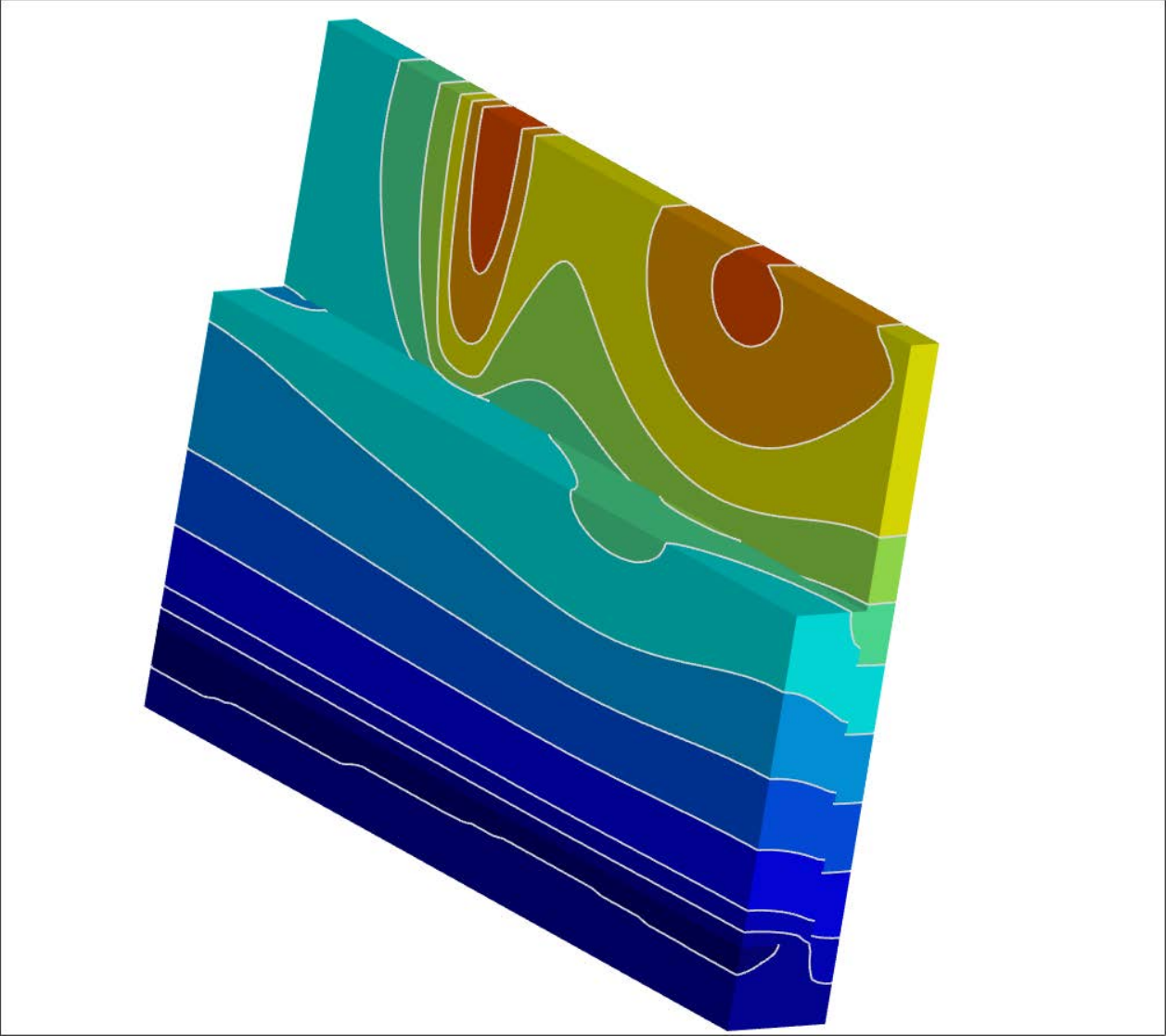}}
\subfloat[$t=0.25t_p$]{\includegraphics[scale=0.17,viewport=50 10 500 530,clip=true]{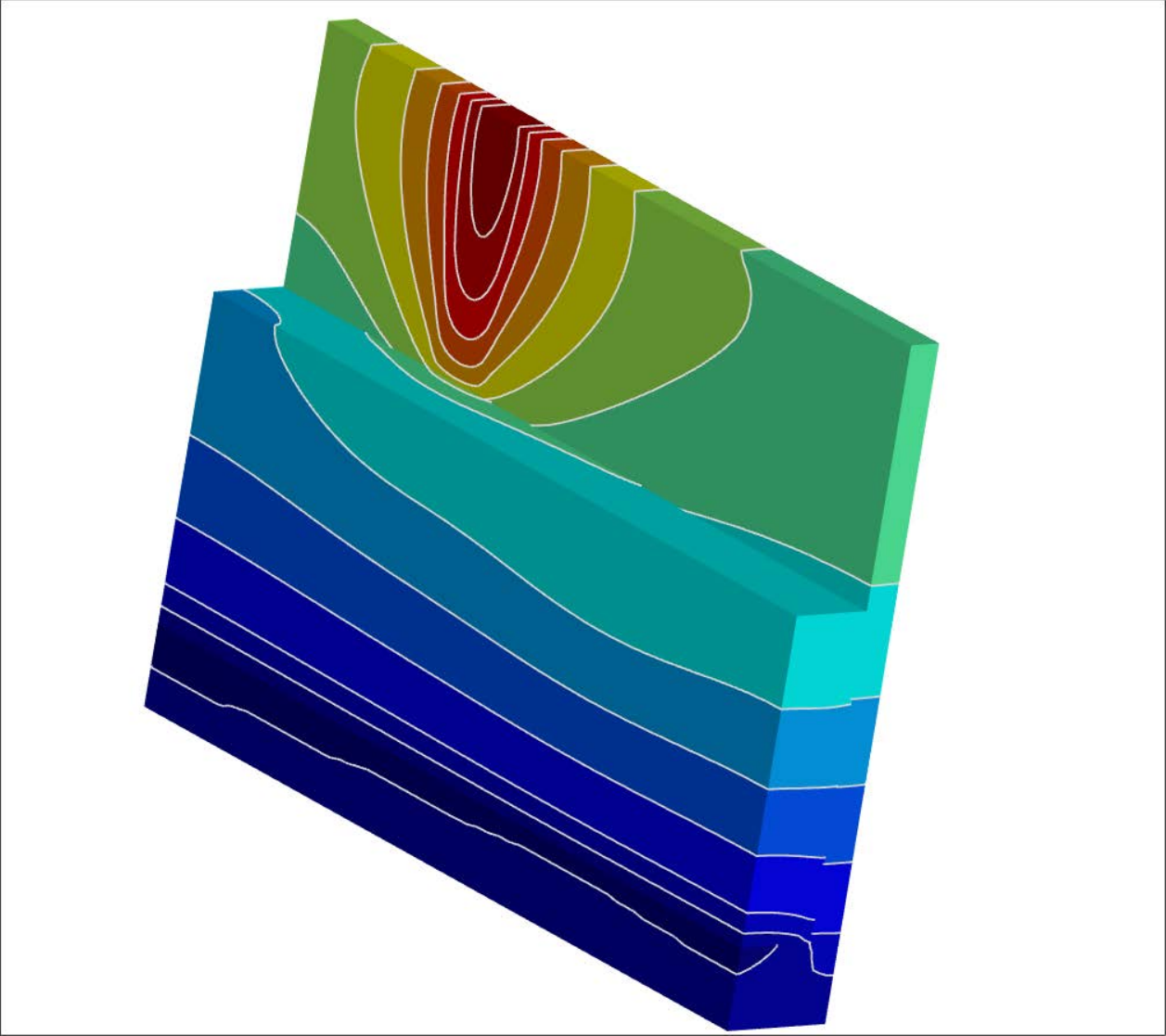}}
\subfloat[$t=0.5t_p$]{\includegraphics[scale=0.17,viewport=50 10 500 530,clip=true]{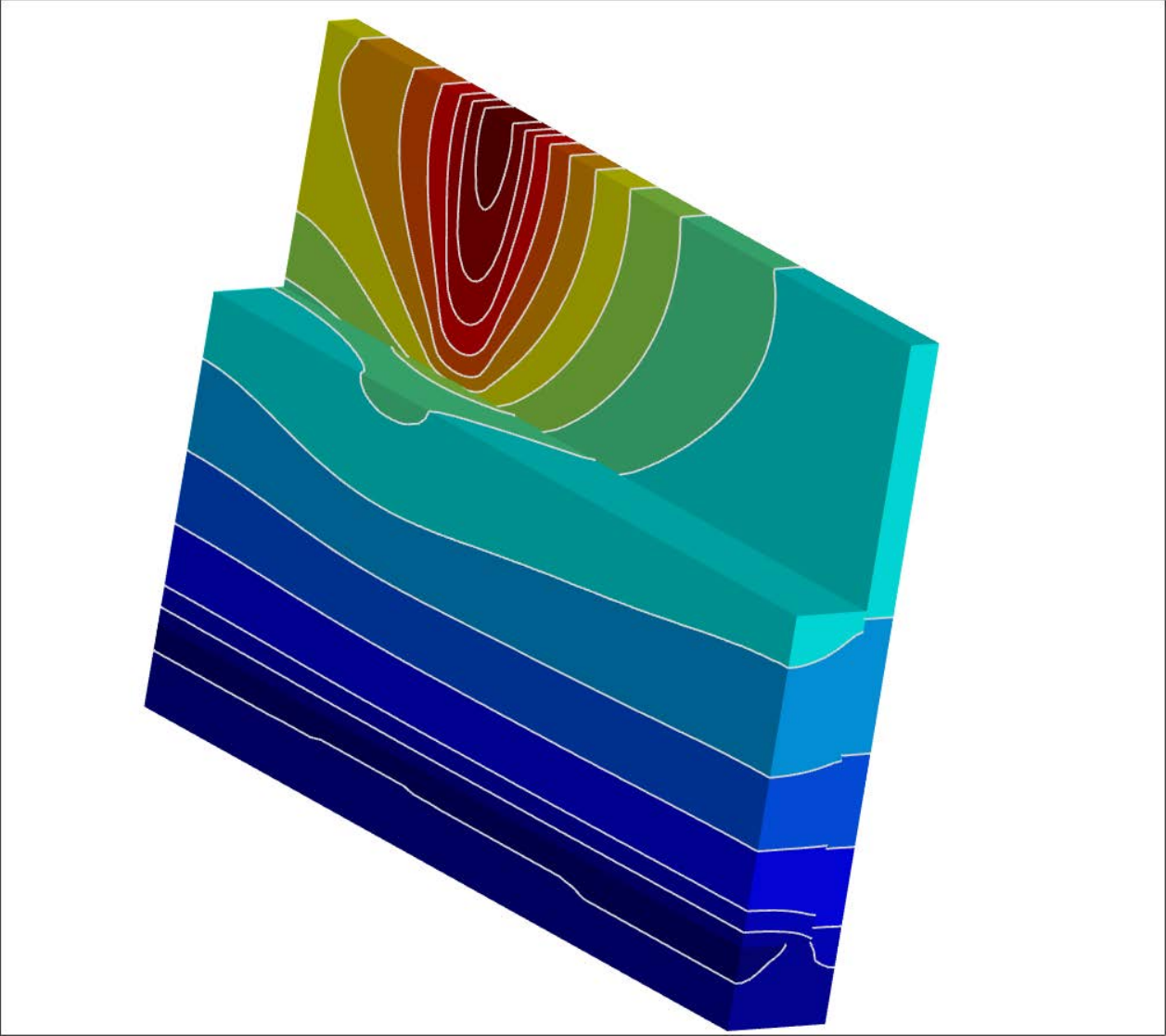}} \\
\subfloat[$t=0.55t_p$]{\includegraphics[scale=0.17,viewport=50 10 500 530,clip=true]{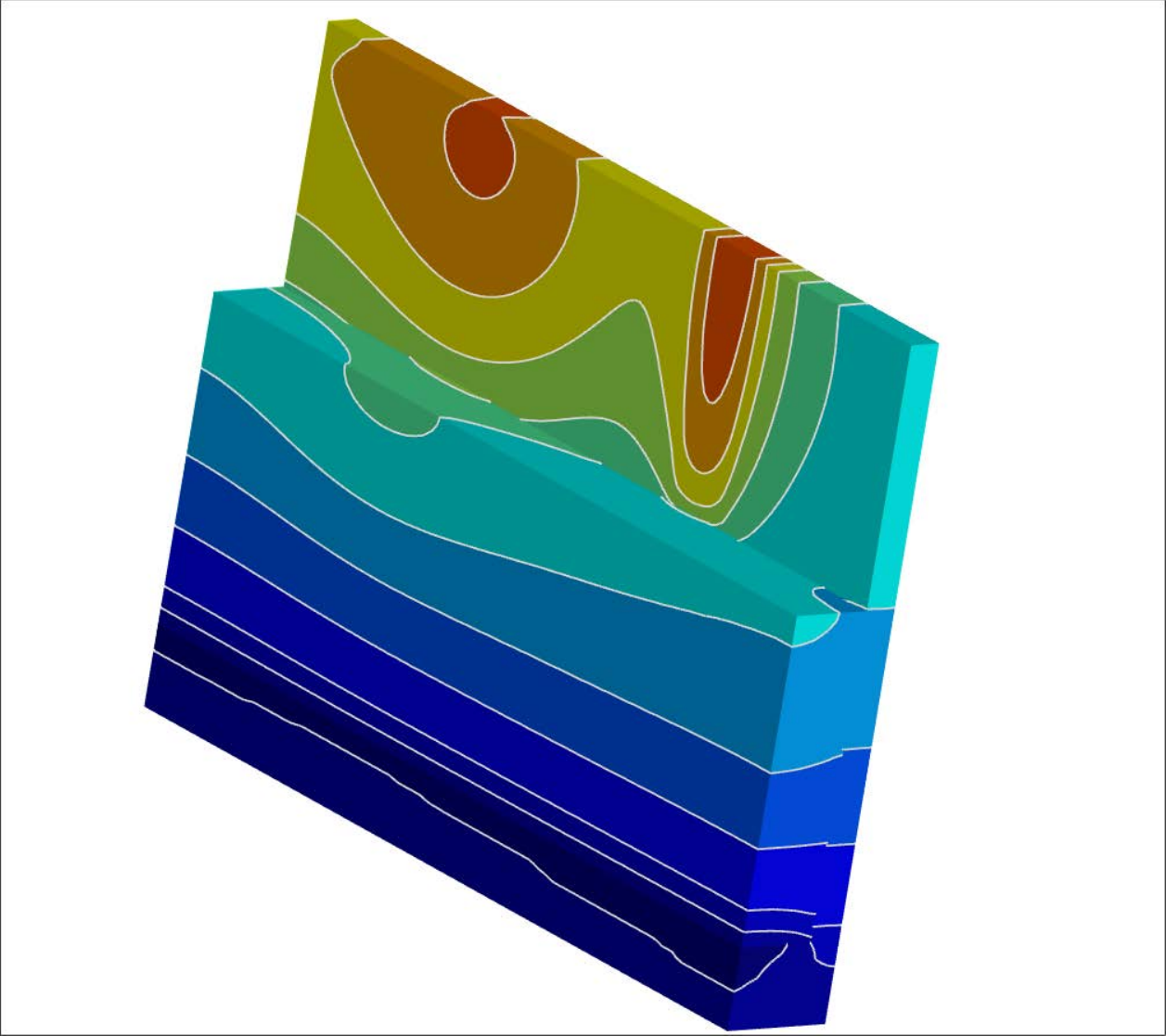}}
\subfloat[$t=0.75t_p$]{\includegraphics[scale=0.17,viewport=50 10 500 530,clip=true]{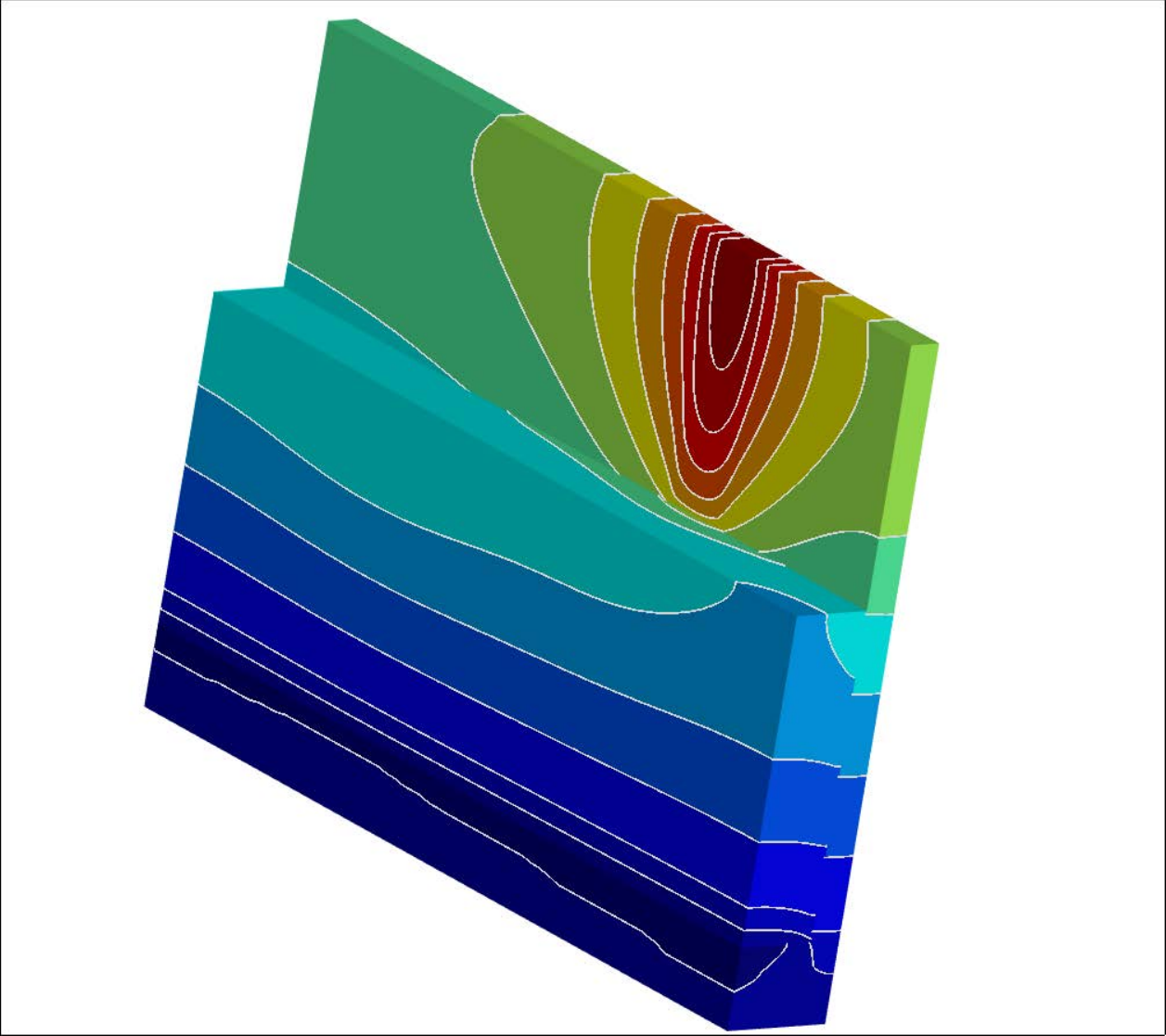}}
\subfloat[$t=1.00t_p$]{\includegraphics[scale=0.17,viewport=50 10 580 530,clip=true]{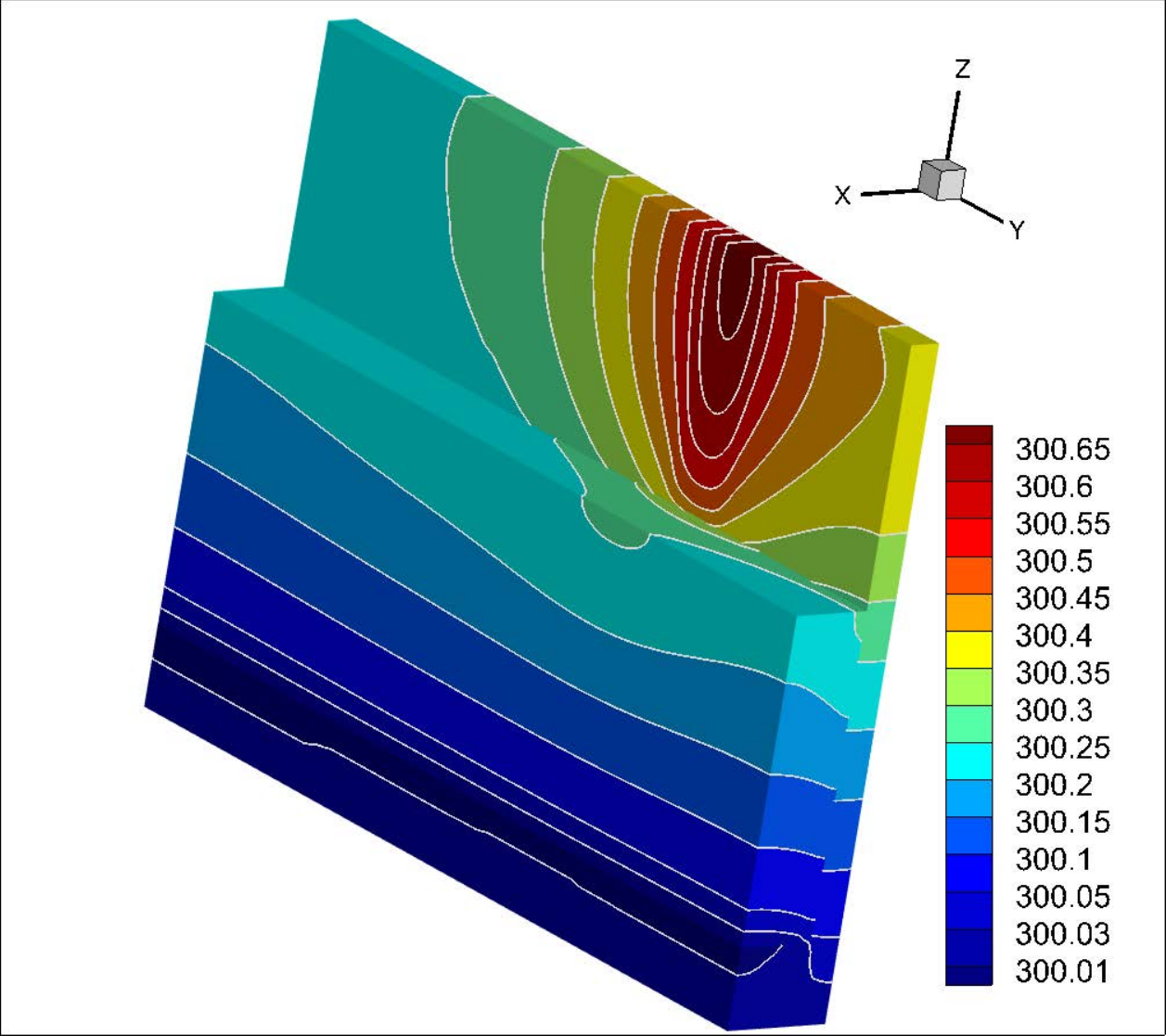}}  
\caption{Temperature contour of bulk FinFET at different moments predicted by BTE under `Alternating' heating when the system reaches the periodic steady state. }
\label{bulkFinFET_two_gate}
\end{figure}
\begin{figure}[htb]
\centering  
\subfloat[$t=0.05t_p$]{\includegraphics[scale=0.17,viewport=50 10 500 530,clip=true]{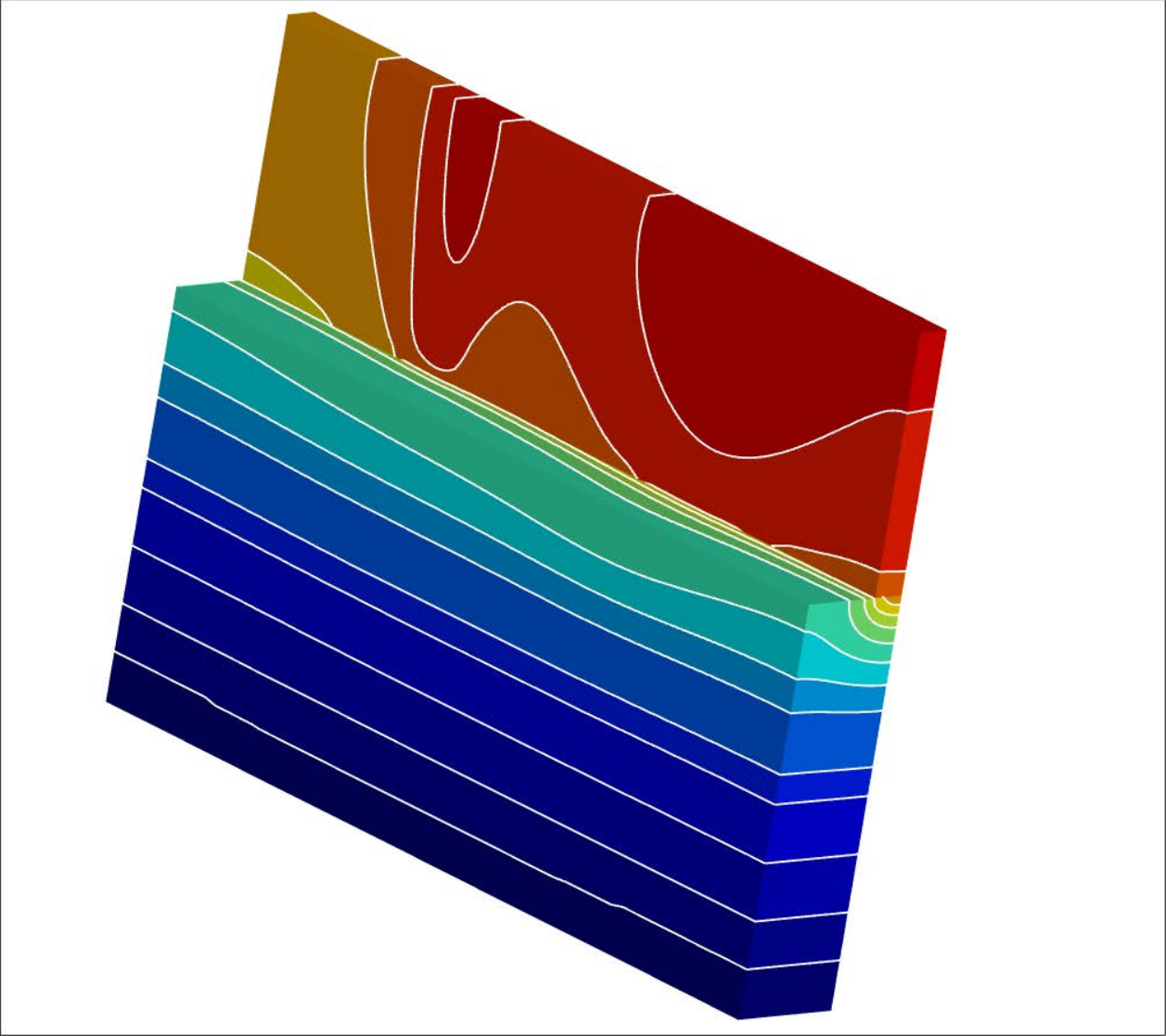}}
\subfloat[$t=0.25t_p$]{\includegraphics[scale=0.17,viewport=50 10 500 530,clip=true]{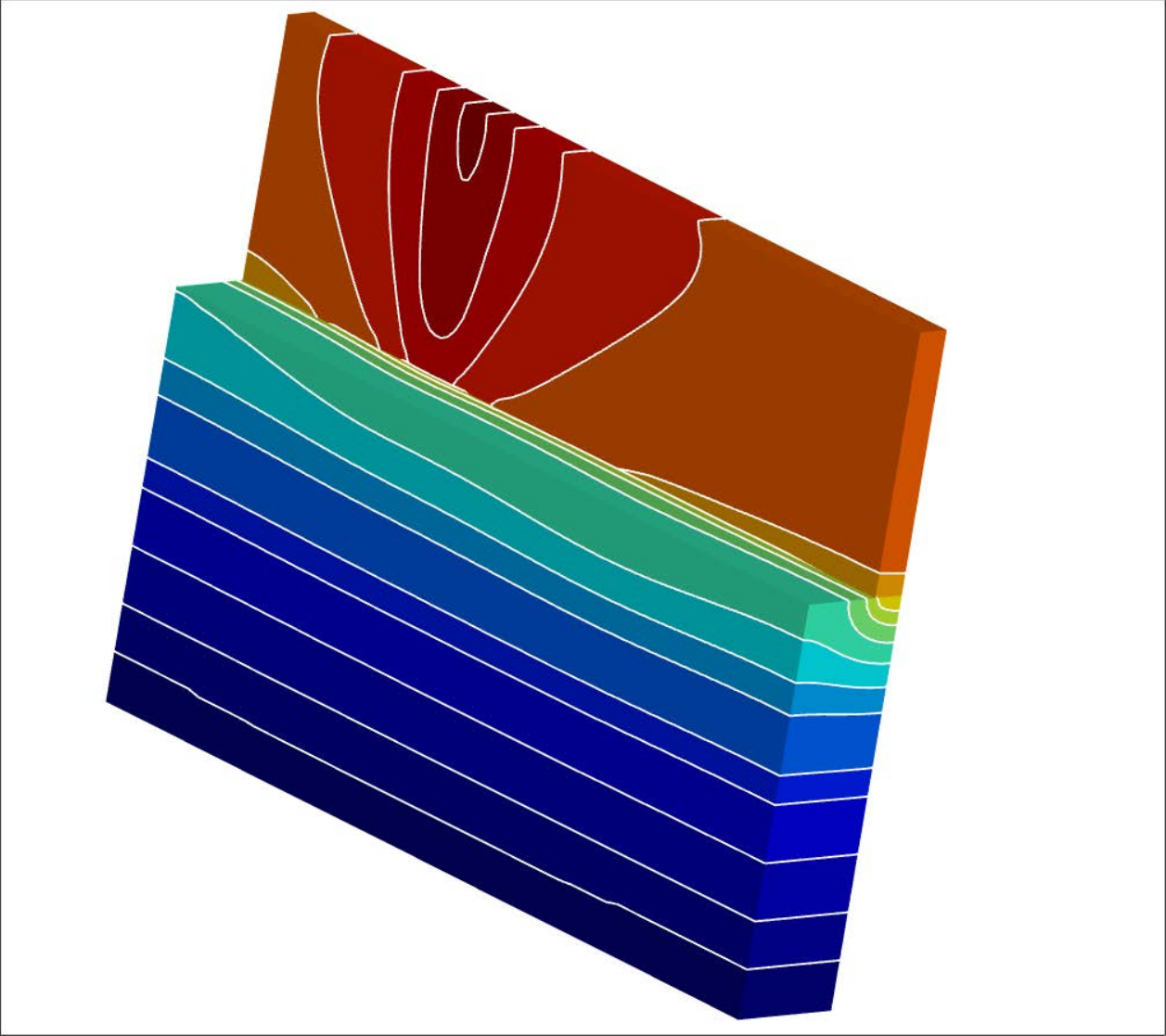}}
\subfloat[$t=0.5t_p$]{\includegraphics[scale=0.17,viewport=50 10 500 530,clip=true]{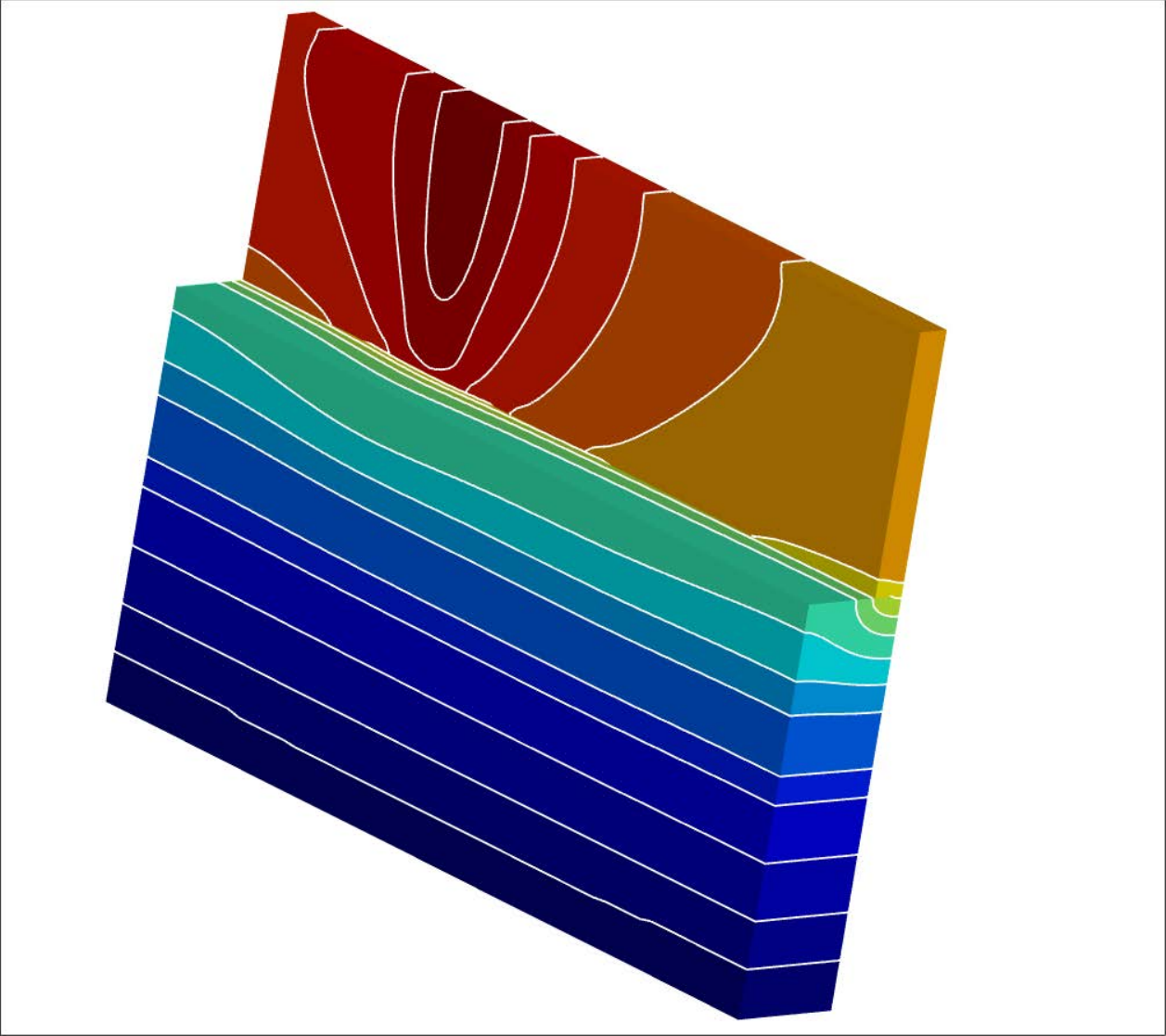}}  \\
\subfloat[$t=0.55t_p$]{\includegraphics[scale=0.17,viewport=50 10 500 530,clip=true]{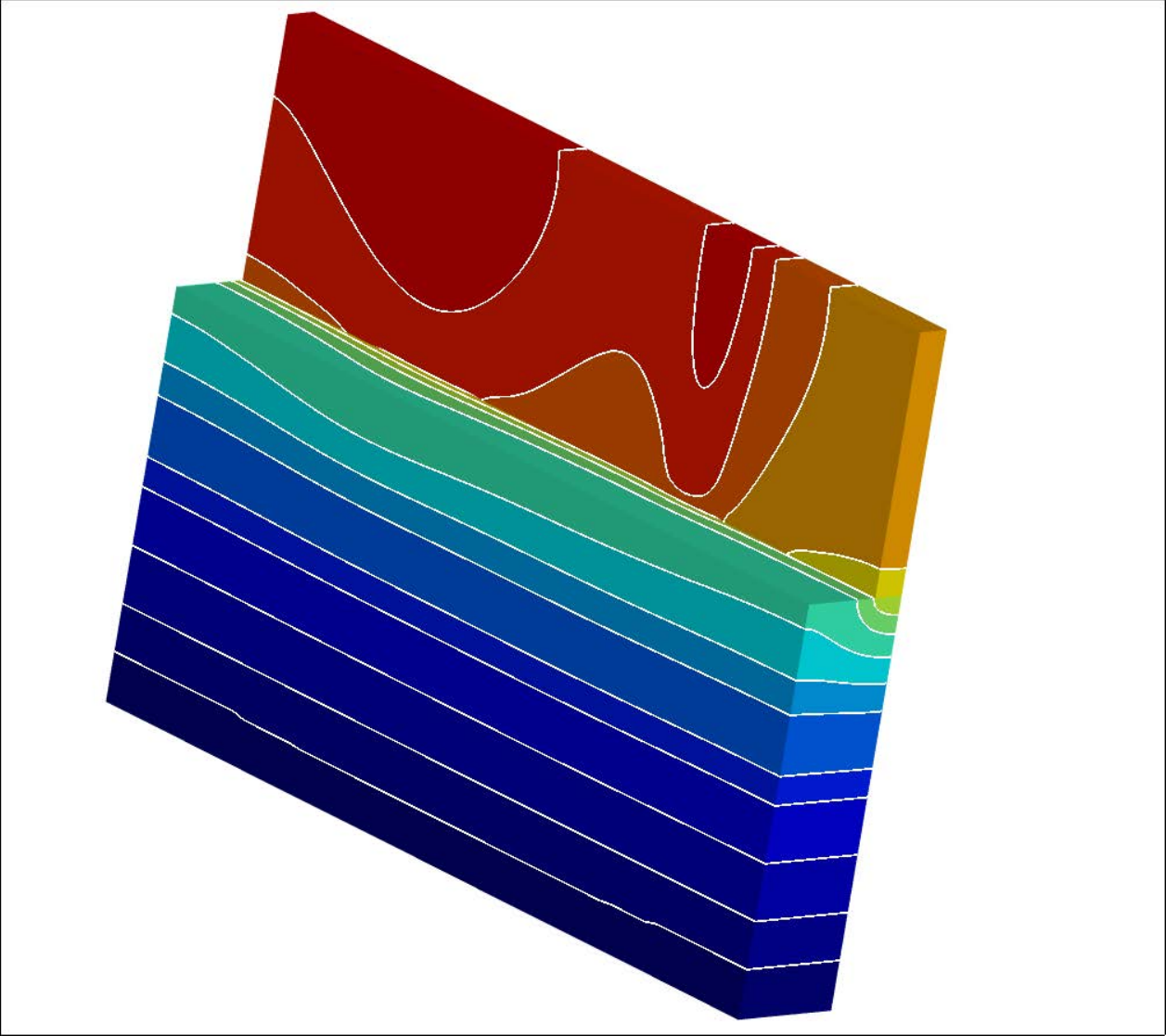}}
\subfloat[$t=0.75t_p$]{\includegraphics[scale=0.17,viewport=50 10 500 530,clip=true]{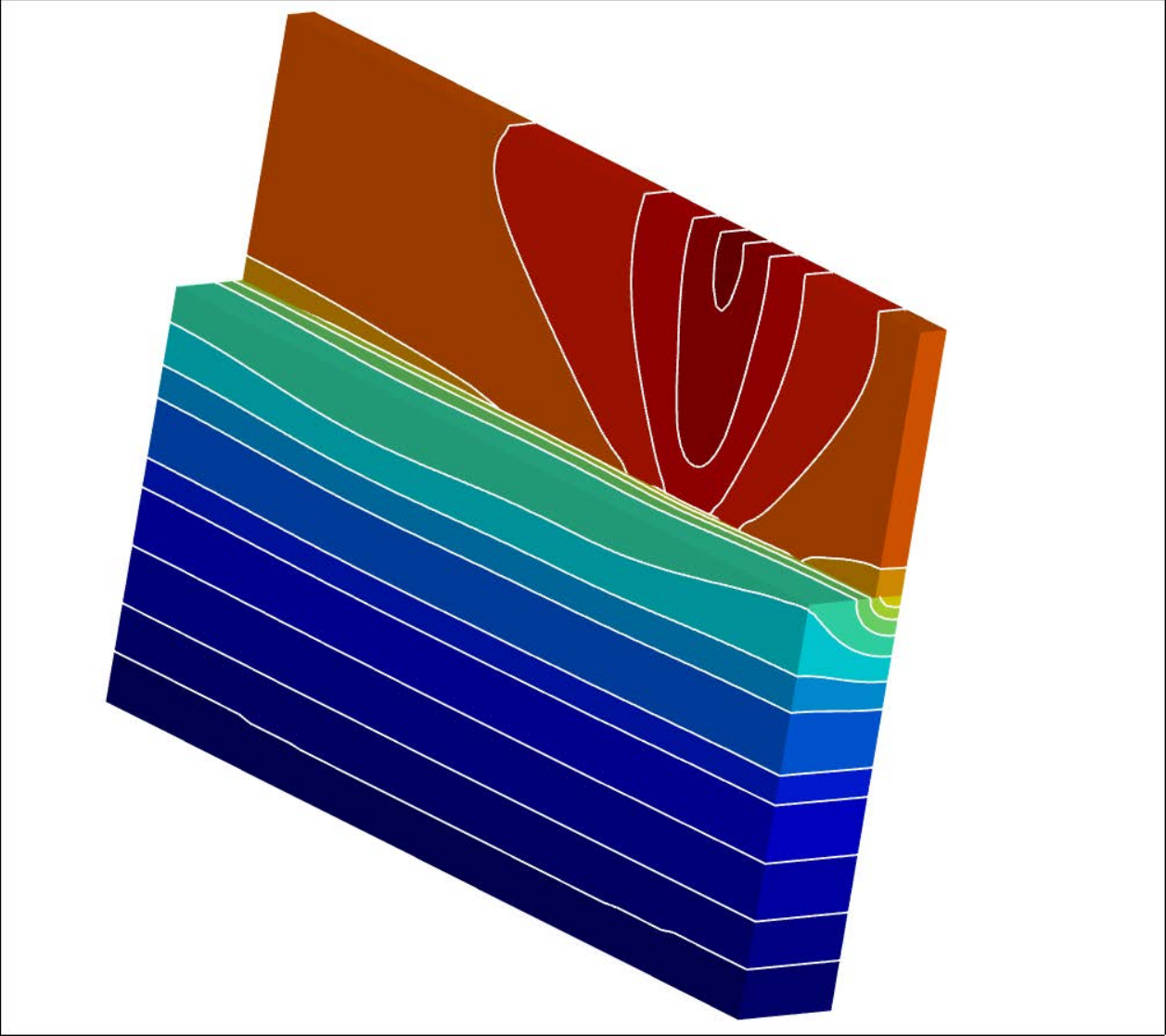}}
\subfloat[$t=1.00t_p$]{\includegraphics[scale=0.17,viewport=50 10 580 530,clip=true]{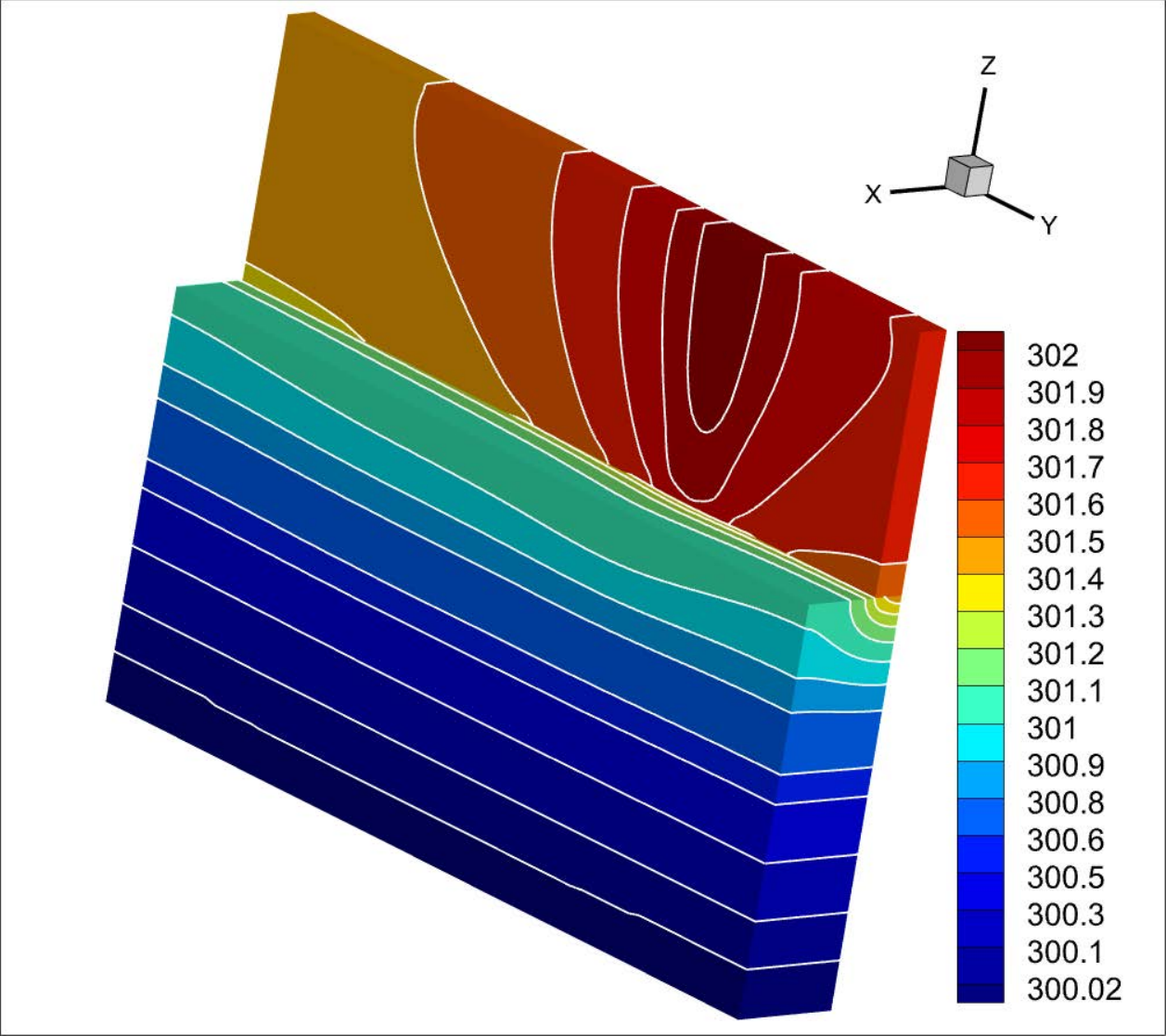}}
\caption{Temperature contour of SOI FinFET at different moments predicted by BTE under `Alternating' heating when the system reaches the periodic steady state.}
\label{SOIFinFET_two_gate}
\end{figure}

Firstly, steady-state temperature fields in bulk or SOI FinFET are simulated and compared in~\cref{3DFinFET_steady} with `continuous heating'.
It can be found that the deviations of temperature fields in the whole domain between the BTE and EFL are smaller in SOI FinFET that those in bulk FinFET.
The underlying physical mechanisms can be explained from the perspective of material composition and phonon transport properties.
Schematics of heat dissipation paths in bulk FinFET and SOI FinFET are shown in~\cref{3DFinFET_steady}(c) and (f), respectively, where there is thermal resistance in the channels between two nodes.
In SOI FinFET~(\cref{3DFinFET_steady}(f)), phonons must pass through two Si/SiO$_2$ interfaces and the silicon dioxide region when they need to transport thermal energy from the heat source region to the heat sink region.
On one hand, the small phonon mean free path in silicon dioxide leads to a diffusive transport, where the EFL is valid.
On the other hand, two Si/SiO$_2$ interfaces significantly increases the thermal resistance and decreases the heat dissipation efficiency.
As shown in~\cref{3DFinFET_steady}(d,e), the temperature gradient is basically along the $z$ direction and the temperature variance presents a linear distribution in the silicon dioxide and bottom silicon substrate areas.
On the contrary, phonons can transfer energy directly from the heat source region to the bottom heat sink only through the silicon material in the bulk FinFET after absorbing a large amount of energy from the heat source, namely, heat dissipation path $1-2-3-6$ in~\cref{3DFinFET_steady}(c). 
In this process, ballistic transport dominates the heat conduction.
Although a lower effective thermal conductivity is introduced in the EFL, the linear assumption between the heat flux and temperature gradient still leads to some deviations compared to the results of phonon BTE.
If we consider the nonlinear, nonlocal, time delay or memory effects in the macroscopic equation, the deviation between the macroscopic simulations and the phonon BTE may be smaller.

In addition, the temperature fields in bulk FinFET predicted by phonon BTE in the bottom silicon substrate areas (\cref{3DFinFET_steady}a) are completely different from those predicted by EFL (\cref{3DFinFET_steady}b).
Actually it is a competition result of two heat dissipation channels with different heat transfer efficiency and various phonon transport behaviors, which can be explained according to the heat dissipation channel $2-4-5$ and $2-3-5$ in~\cref{3DFinFET_steady}(c).
In heat dissipation channel $2-4-5$, on one hand, two interfacial thermal resistance exists in the thermal transport process from $2$ to $4$ and from $4$ to $5$.
One the other hand, the phonon mean free path of silicon dioxide is much smaller than $8$ nm.
In heat dissipation channel $2-3-5$, the transfer of thermal energy only happens in the silicon materials without interfacial thermal resistance and ballistic phonon transport dominates heat conduction.
Compared tp channel $2-3-5$, there is larger thermal resistance and lower phonon transport efficiency in channel $2-4-5$.
Hence, it can be found that the temperature contour line in bottom silicon areas is even perpendicular to the bottom surface.
When using the EFL, ballistic phonon transport in heat dissipation channel $2-3-5$ is replaced by the temperature diffusion with an effective thermal conductivity.
It is well known that one of the drawbacks of the diffusion equation is that it has an infinite heat propagation speed~\cite{RevModPhysJoseph89}, namely, the effect of a temperature fluctuation or change at any spatial point can instantly affect the entire region.
Hence, it can be found that the temperature contour line mainly parallel to the bottom surface.

Secondly, the evolution of transient peak temperature over time in bulk or SOI FinFET under three heating strategies is shown in~\cref{3DFinFET_transient}.
The temperature predicted by EFL rises more slowly than BTE at the initial stage, and it also takes longer to reach the periodic steady state.
When the system reaches the periodic steady state, the transient temperature predicted by EFL is also inconsistent with the BTE results.
For example, the `Intermittent' heating has a higher peak temperature rise than that of `Alternating' heating in the heating stage and has a lower peak temperature in the second half of the heating period in the BTE solutions.
However, in the EFL prediction results, the maximum temperature of the `Alternating' heating is lower than that of the `Intermittent' heating in the second half of the heating period.
Compared to `Intermittent' heating, the temperature variance of `Alternating' heating is smaller.

Thirdly, the transient heat dissipation process in bulk or SOI FinFET under `Intermittent' or `Alternating' heating is described.
The temperature contour predicted by BTE at different moments is plotted when the system reaches the periodic steady state in~\cref{bulkFinFET_transient,SOIFinFET_transient,bulkFinFET_two_gate,SOIFinFET_two_gate}, where $t=0.25t_p$ represents that the current moment is a quarter of a heating period.
The periodic steady state represents that the temporal and spatial distributions of the temperature field in the current heating period is the same as those in the next heating period, with a relative deviation of less than $10^{-6}$.

In bulk FinFET under `Intermittent' heating, when the external heat source begins to heat the system, e.g., $t=0.05t_p$, the temperature near the heat source areas begins to rise continuously, and the high thermal energy in the fin region is rapidly transferred from hotspot area to other geometric regions by phonon ballistic transport. 
At the contact interface between the silicon fin and the silicon dioxide insulation layer, the temperature in the silicon region is higher than that of the silicon dioxide region at the same $z$ height.
However, when $t=0.25t_p$ or $t=0.5t_p$, the temperature in the silicon region is lower than that of the silicon dioxide region at the same $z$ height.
This is because the mean free path or thermal diffusivity rate of silicon is much higher than that of silicon dioxide, which leads to a higher heat dissipation efficiency. 
The thermal energy in silicon dioxide is mainly transferred from silicon to silicon dioxide through the interface, and then transferred to the bottom heat sink. 
As shown in~\cref{3DFinFET_steady}(c), the heat dissipation efficiency of channel $2-3-6$ is much higher than that of $2-4-5-6$. 
Although the thermal energy is transferred from silicon to silicon dioxide through the interface at the beginning $t=0.05t_p$, the energy in silicon dioxide is not efficiently transferred to the bottom heat sink due to its small mean free path, small thermal diffusivity and large interfacial thermal resistance.
It is similar to a thermal reservoir. 
Over time, the silicon dioxide region actually got hotter, even higher than that of silicon at the same $z$ height.
When the heat source is removed $t>0.5t_p$, the peak temperature decreases significantly due to the large mean free path of silicon.

In SOI FinFET under `Intermittent' heating, it is well known that the silicon dioxide insulation layer is mainly introduced to improve electrical performance in the actual chip design~\cite{IRDS2023,heat_chip_2019}, but inevitably, it increases the temperature of silicon fin area due to low thermal conductivity.
From~\cref{SOIFinFET_transient}, it can be observed that the temperature of the silicon fin region changes dramatically with time in a heating period, which actually generates a large thermal shock to the materials.
Oppositely, the temperature contour profiles in the silicon dioxide insulation layer and silicon substrate region are basically flat, showing a linear distribution.
In other words, the silicon dioxide insulation layer reduces the thermal shock on the bottom substrate material although it  raised the overall temperature in the fin area.

For `Alternating' heating, two external heat source heat the system in turn so that the thermal conduction characteristics in the $y$ direction are no longer symmetrical.
Temperature contour in bulk or SOI FinFET at different moments under `Alternating' heating when the system reaches the periodic steady state is plotted in~\cref{bulkFinFET_two_gate,SOIFinFET_two_gate}.
When $t \leq 0.5t_p$, one heat source starts to provide thermal energy so that the hotspot temperature increases, while the other is removed so that the hotspot temperature decreases gradually.
Therefore, compared to the `Intermittent' heating in~\cref{bulkFinFET_transient,SOIFinFET_transient}, we can see that there is always a high temperature hotspot in the silicon fin region.
This actually reflects that the overall temperature in the silicon fin region under `Alternating' heating fluctuates less over time than that of under `Alternating' heating, which is also verified in~\cref{3DFinFET_transient}.
Less temperature fluctuations represents smaller thermal shock on materials, which could delay the material life to some extent.

\section{CONCLUSION}
\label{sec:conclusion}

Steady/unsteady heat dissipation in nanoscale bulk or SOI transistors under different heating strategies is investigated by the phonon BTE.
Results show that it is not easy to accurately capture the heat conduction in transistors by the EFL although the effect of boundary scattering on phonon transport is added into the effective thermal conductivity.
There are still some deviations between the results of phonon BTE and EFL, especially near the hotspot areas where ballistic phonon transport dominates and the temperature diffusion is no longer valid.
Although the silicon dioxide increases the peak temperature significantly, it makes the temperature profiles in the silicon dioxide insulation layer and silicon substrate region flat, which reduces the dramatic temperature fluctuations.
Different heating strategies have great influence on the peak temperature rise and transient thermal dissipation process.
Compared to `Intermittent' or `Continuous' heating, the temperature variance of `Alternating' heating is smaller, which indicates that this heating strategy also reduces the dramatic temperature fluctuations.

\section*{Acknowledgment}
Q.L. acknowledges the support of the National Natural Science Foundation of China (52376068).
C.Z. acknowledges the members of online WeChat Group: Device Simulation Happy Exchange Group, for extensive discussions.
The authors acknowledge Beijng PARATERA Tech CO.,Ltd. for providing HPC resources that have contributed to the research results reported within this paper.

\appendix

\section{Discrete unified gas kinetic scheme}
\label{sec:dugks}

The interfacial thermal resistance $R_{eff}$ between two dissimilar solid materials is~\cite{RevModPhys.94.025002,dengke_2024_APL}
\begin{align}
R_{eff} = \frac{\delta T}{q}, 
\end{align}
where $\delta T$ and $q$ are the temperature drop between the two sides of the interface and the heat flux across the interface, respectively.
In order to deal with the interfacial thermal resistance between silicon and silicon dioxide materials, the diffuse mismatch model is used~\cite{dengke_2024_APL,RevModPhys.94.025002,JAP_qinghao_2017,IEEE_qinghao_2018}, which assumes that all phonons loses previous memories and completely follows the diffuse transmitting or reflecting rule after interacting with the interface.
Transmittance and reflectance on each side of the interface satisfy 
\begin{align}
r_{12} + t_{12} &=1
\end{align}
due to energy conservation, where $t_{12}$ (or $t_{21}$) represents the transmittance from medium $1$ (or medium $2$) to medium $2$ (or medium $1$) across the interface, and $r_{12}$ (or $r_{21}$) represents the reflectance in the medium $1$ (or medium $2$) reflected back from the interface.
Note that the net heat flux across the interface should be zero at the thermal equilibrium state due to the principle of detailed balance so that
\begin{align}
t_{12} C_1 v_1  =  t_{21} C_2 v_2,
\end{align}
where $C_1$ and $v_1$ (or $C_2$ and $v_2$) are the specific heat and group velocity of medium $1$ (or medium $2$).

\begin{figure}[htb]
\centering  
\subfloat[]{\includegraphics[width=0.28\textwidth]{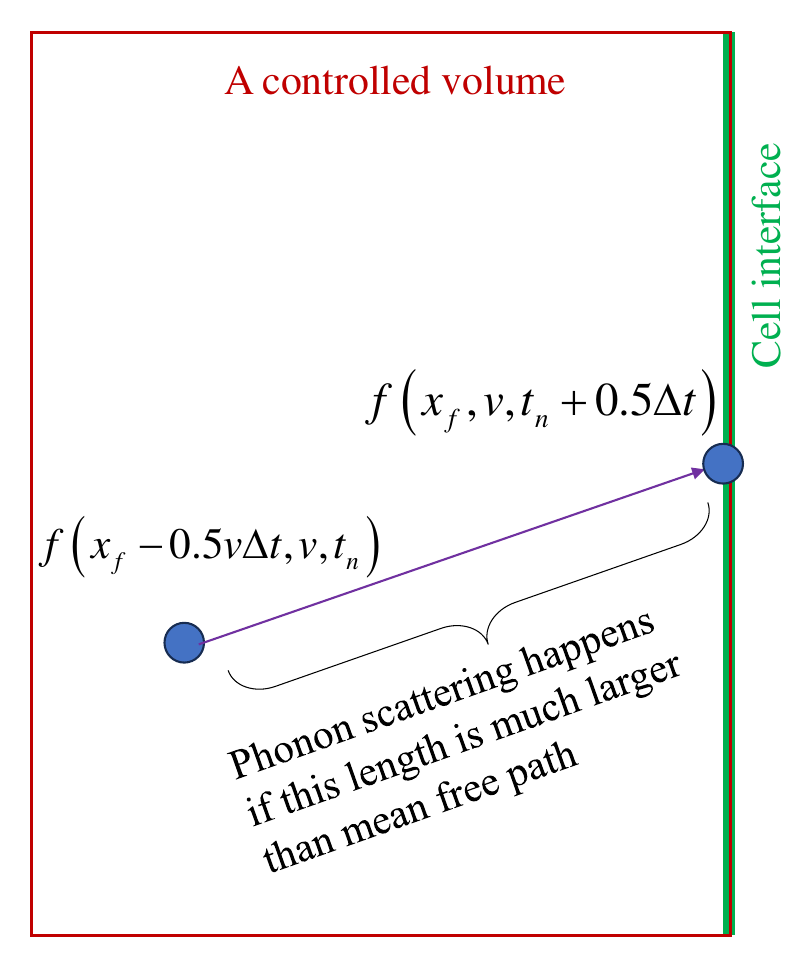}}
\subfloat[]{\includegraphics[width=0.18\textwidth]{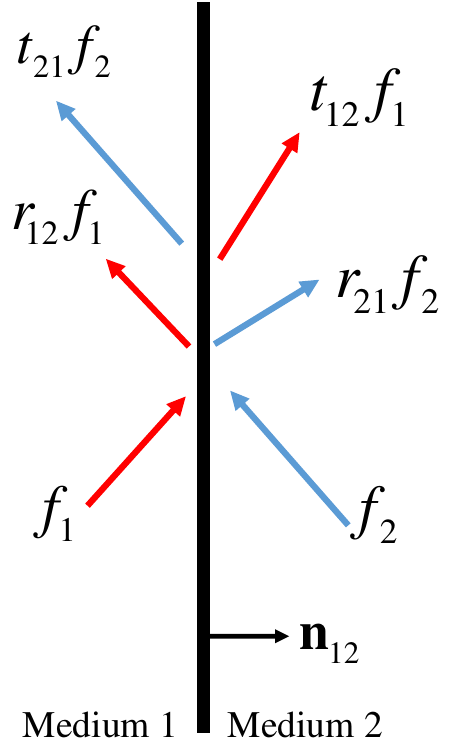}}  \\
\subfloat[]{\includegraphics[width=0.4\textwidth]{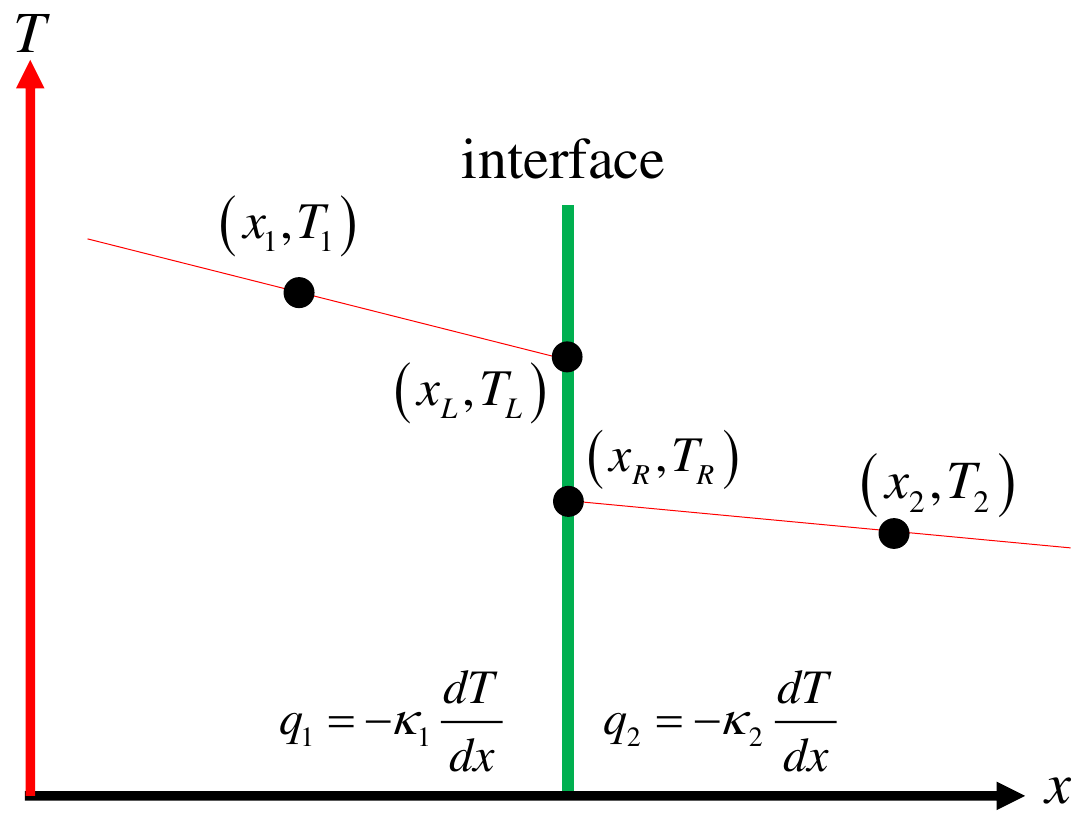} } 
\caption{(a) Reconstruction of phonon distribution function at the cell interface for a given discretized controlled volume. (b) Transmittance and reflectance of phonon distribution function on each side of the interface between medium $1$ and medium $2$. (c) Schematic of the spatial distributions of temperature near the interface between medium $1$ and $2$.
}
\label{BTEdmm}
\end{figure}
Discrete unified gas kinetic scheme (DUGKS) is used to solve the transient frequency-independent phonon BTE.
More specific numerical solution process and boundary treatments can be found in previous papers~\cite{GuoZl16DUGKS,guo_progress_DUGKS}.
Here we make a brief introduction of DUGKS.
Under the discretized six-dimensional phase space, the phonon BTE is
\begin{align}
&\frac{f_{i,k}^{n+1} - f_{i,k}^{n}}{\Delta t} + \frac{\Delta t}{V_i} \sum_{j \in N(i)} \left(  v_g \bm{s}_{k}  \cdot \mathbf{n}_{ij} f_{ij,k}^{n+1/2} S_{ij} \right)  \notag \\
&= \frac{\Delta t}{2}  \left( H_{i,k}^{n+1}  +H_{i,k}^{n} \right),
\label{eq:dugkscenter}
\end{align}
where $i$, $k$, $ij$, $n$ are the indexes of cell center, solid angle, cell interface and time step at the finite-volume discrete level, respectively.  $H = (f^{eq } -f  )/ \tau  + \dot{S} /(4 \pi) $. 
In order to update the phonon distribution function from $f_{i,k}^{n}$ to $f_{i,k}^{n+1}$, the key is the reconstruction of the distribution function at the cell interface at the half time step $f_{ij,k}^{n+1/2}$.
When a phonon is transferred from the inner region of a control volume to the interface after half a time step, it will suffer a large number of phonon scattering processes if the time step is much longer than the phonon relaxation time or this length is much larger than mean free path, as shown in~\cref{BTEdmm}(a).
To respect this physical law, the phonon BTE is solved again along the direction of group velocity and trapezoidal quadrature is used for the time integration of the scattering terms,
\begin{align}
&f(\bm{x}_f, \bm{v}, t_{n}+ 0.5 \Delta t )- f(\bm{x}_1, \bm{v}, t_{n} )  \notag \\
&=\frac{\Delta t}{4}  \left(   H(\bm{x}_f,t_{n}+ 0.5\Delta t)+H(\bm{x}_1,t_{n})  \right)   \label{eq:dugksface}  \\
\Longrightarrow  & \bar{f}(\bm{x}_f, \bm{v}, t_{n}+ 0.5 \Delta t ) =  \bar{f}^{+}(\bm{x}_1, \bm{v}, t_{n} ),
\end{align}
where $\bm{x}_f$ is the center of cell interface, $\bm{x}_1=\bm{x}_f-0.5\bm{v}\Delta t$, $\bar{f}=f-\frac{\Delta t}{4} H $, $\bar{f}^{+}= f+\frac{\Delta t}{4} H$.
Equations~\eqref{eq:dugkscenter} and~\eqref{eq:dugksface} are the key evolution process of DUGKS~\cite{GuoZl16DUGKS}, that is, solving the discrete BTE at the cell center in a complete time step, while coupling phonon advection and scattering together in the reconstruction of the interfacial distribution function at the half time step, through which it allows the time step or cell size to be much larger than the relaxation time and phonon mean free path in the (near) diffusive regime.
The phonon distribution function incident into the cell interface is
\begin{align}
f(\bm{x}_f)= \frac{4 \tau}{4 \tau +\Delta t} \left( \bar{f}(\bm{x}_f)  + \dot{S}/(4 \pi) + \frac{\Delta t}{4 \tau} f^{eq}(\bm{x}_f)  \right),
\label{eq:fbar}
\end{align}
from which it can be found that we have to firstly calculate the equilibrium state at the cell interface $f^{eq}(\bm{x}_f)$ if we want to reconstruct the distribution function $f(\bm{x}_{f})$ based on $\bar{f}(\bm{x}_f)$.

Here we focus on how to reconstruct the transmission and reflection distribution function at the interface between two dissimilar solid materials.
As shown in~\cref{BTEdmm}(b), the phonon transmission and reflection at the interface are related to the phonon distribution on both sides.
Diffuse mismatch model assumes that the phonon distribution at the interface pointing from interface to medium $1$ (or medium $2$) follows the equilibrium distribution with temperature $T_1^p$ (or $T_2^p$).
Then we have
\begin{align}
&\int_{ \bm{v}_1 \cdot \mathbf{n}_{12} > 0 }  - r_{12}  f_1  \bm{v}_1 \cdot \mathbf{n}_{12}  d \Omega +  \int_{ \bm{v}_2 \cdot \mathbf{n}_{12} < 0 } t_{21}  f_2 \bm{v}_2 \cdot \mathbf{n}_{12}  d \Omega \notag \\
 &= \int_{ \bm{v}_1 \cdot \mathbf{n}_{12} < 0 }  f^{eq} (T_1^p ) \bm{v}_1 \cdot \mathbf{n}_{12}  d \Omega  \label{eq:dmmflux1} \\
&\int_{ \bm{v}_1 \cdot \mathbf{n}_{12} > 0 }  t_{12}  f_1  \bm{v}_1 \cdot \mathbf{n}_{12}  d \Omega +  \int_{ \bm{v}_2 \cdot \mathbf{n}_{12} < 0 } -r_{21}  f_2 \bm{v}_2 \cdot \mathbf{n}_{12}  d \Omega \notag \\
 &= \int_{ \bm{v}_2 \cdot \mathbf{n}_{12} > 0 }  f^{eq} (T_2^p ) \bm{v}_2 \cdot \mathbf{n}_{12}  d \Omega  \label{eq:dmmflux2} \\
&\int_{ \bm{v}_1 \cdot \mathbf{n}_{12} > 0 } f_1  d \Omega +  \int_{ \bm{v}_1 \cdot \mathbf{n}_{12} < 0 } f^{eq} (T_1^p) d \Omega = \int f^{eq} (T_1 )  d \Omega, \label{eq:dmmT1} \\
&\int_{ \bm{v}_2 \cdot \mathbf{n}_{12} < 0 } f_2  d \Omega +  \int_{ \bm{v}_2 \cdot \mathbf{n}_{12} > 0 } f^{eq} (T_2^p) d \Omega = \int f^{eq} (T_2 )  d \Omega,  \label{eq:dmmT2} \\
&f_1   =   \frac{4 \tau_1}{4 \tau_1 +\Delta t} \left( \bar{f}_1  +  \dot{S}/(4 \pi) + \frac{\Delta t}{4 \tau_1} f^{eq}(T_1)\right), \quad \bm{v}_1 \cdot \mathbf{n}_{12} > 0  \\
&f_2  =   \frac{4 \tau_2}{4 \tau_2 +\Delta t} \left( \bar{f}_2  +  \dot{S}/(4 \pi) + \frac{\Delta t}{4 \tau_2} f^{eq}(T_2)\right), \quad \bm{v}_2 \cdot \mathbf{n}_{12} < 0  
\end{align}
where $\mathbf{n}_{12}$ is the unit normal vector pointing from media $1$ to media $2$, $T_1$ and $T_2$ are the local equivalent equilibrium temperatures on each side of the interface. 
$\tau_1$ (or $\tau_2$) and $\bm{v}_1$ (or $\bm{v}_2$) are the phonon properties in medium $1$ (or medium $2$).
The first two equations (\ref{eq:dmmflux1},\ref{eq:dmmflux2}) come from the physical assumptions of diffuse mismatch model and the conservation of heat flux across the interface, for example, in Eq.~\eqref{eq:dmmflux1}, all phonons in medium $1$ emitted from the interface, including the phonons in medium $1$ reflecting back from the interface and the phonons in medium $2$ transmitting across the interface, follow the equilibrium distribution with temperature $T_1^p$.  
$T_1$ (\ref{eq:dmmT1}) and $T_2$ (\ref{eq:dmmT2}) are calculated by taking the moment of distribution function over the whole momentum space.
Combined above six equations, $T_1$, $T_2$, $T_1^p$ and $T_2^p$ can be obtained by Newton method.

\section{Numerical discretizations and independence test}
\label{sec:discretization}

To solve the transient phonon BTE, the temporal space, solid angle space and spatial space are both discretized into a lot of small pieces.
Time step is $\Delta t=0.125$ ps and Cartesian grids with uniform cell size $\Delta x=1$ nm is used.
Solid angle $\bm{s}=\left( \cos \theta, \sin \theta \cos \varphi, \sin \theta  \sin \varphi  \right)$ is discretized into $N_{\theta}\times N_{\varphi} = 48 \times 48$ pieces, where $\cos \theta \in [-1,1]$ is discretized by $N_{\theta}$-point Gauss-Legendre quadrature and $\varphi \in [0,\pi]$ (due to symmetry) is discretized by the  Gauss-Legendre quadrature with $N_{\varphi}/2$ points.
All numerical results are obtained by a three-dimensional C/C++ program.
MPI parallelization computation with $48$ CPU cores based on the decomposition of solid angle space is implemented and the discrete solid angles corresponding to the specular reflection are ensured in the same CPU core.
\begin{figure}[htb]
\centering  
\subfloat[3D, temperature]{\includegraphics[width=0.22\textwidth]{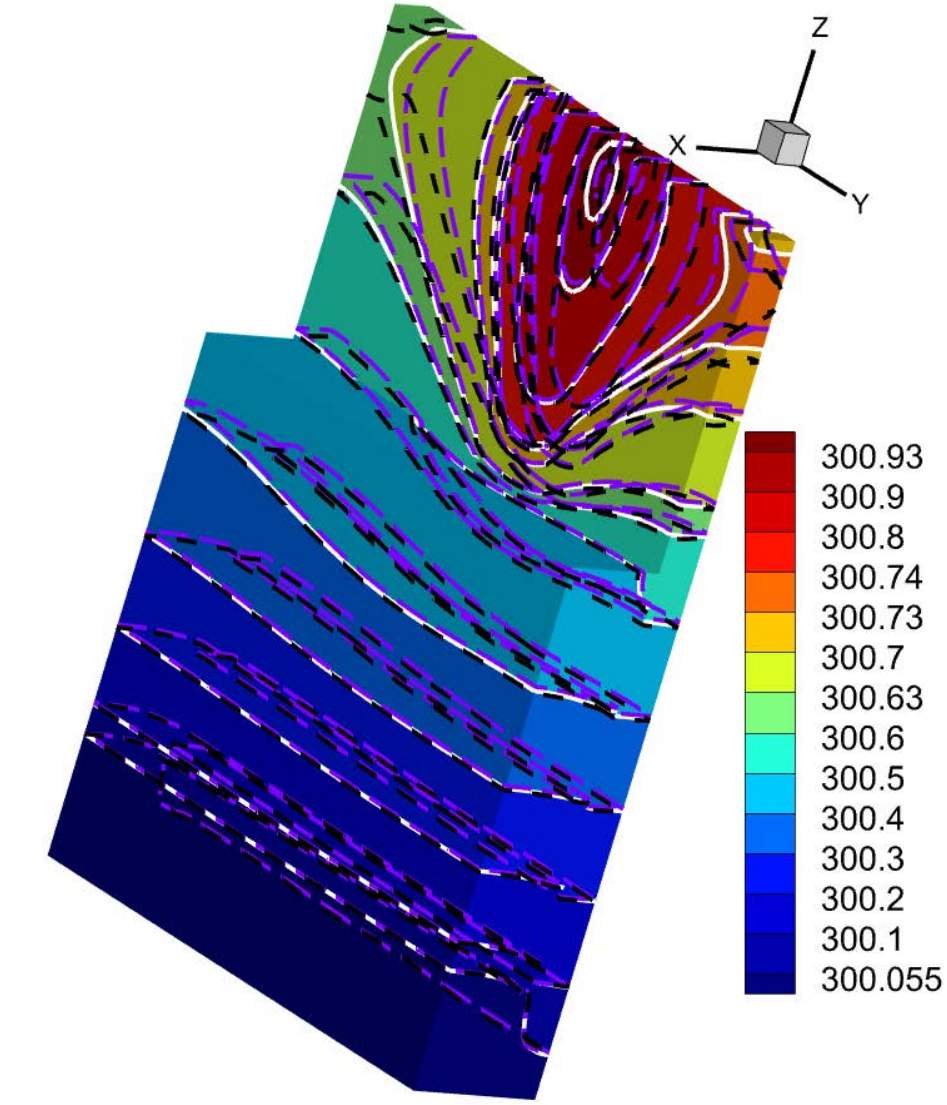}}  ~~  
\subfloat[3D YZ slice, temperature]{\includegraphics[width=0.22\textwidth]{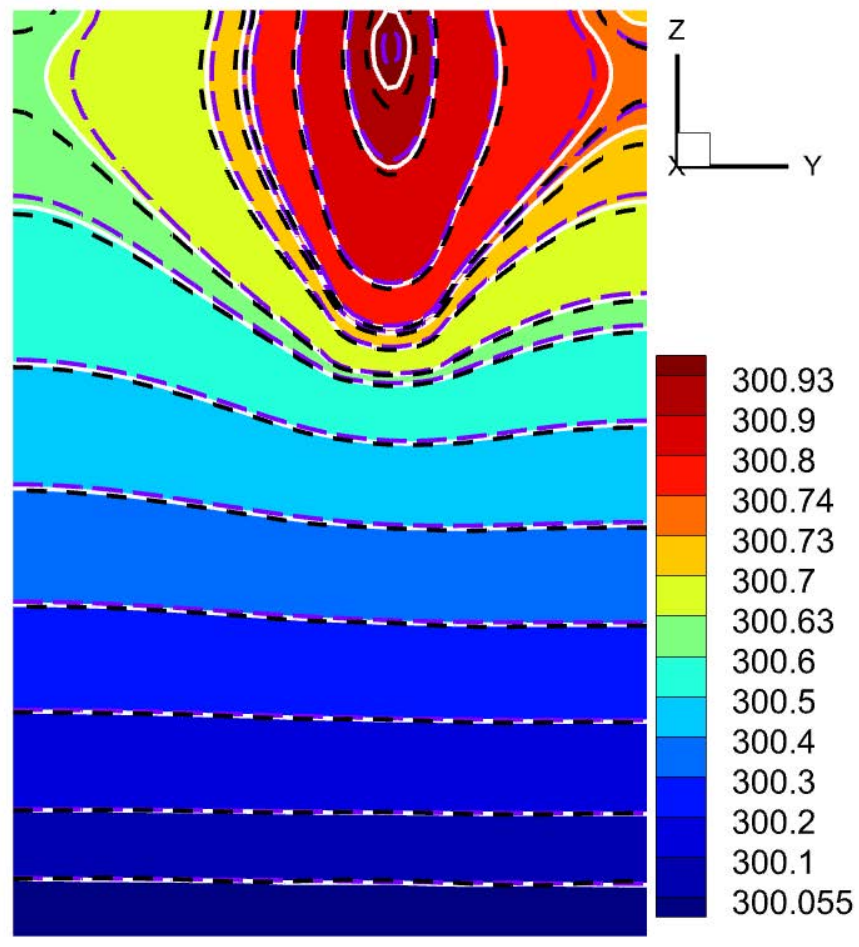}} \\
\subfloat[3D YZ slice, heat flux]{\includegraphics[width=0.22\textwidth]{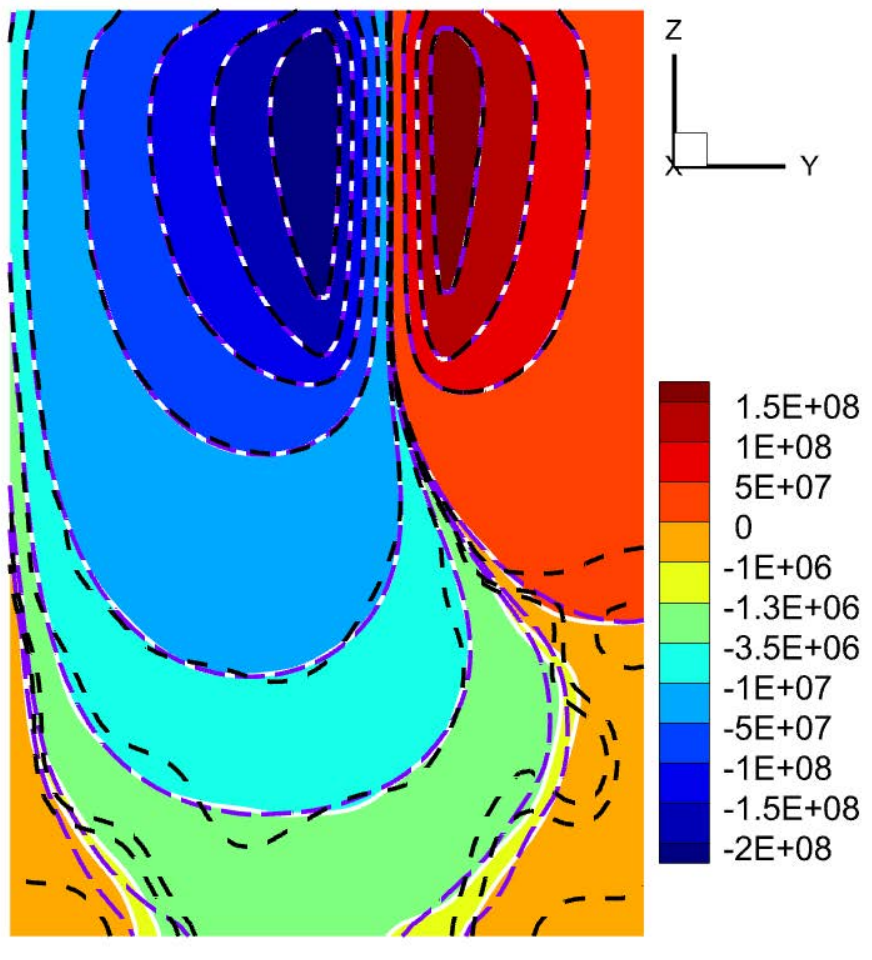} } ~~~
\subfloat[3D YZ slice, heat flux]{\includegraphics[width=0.22\textwidth]{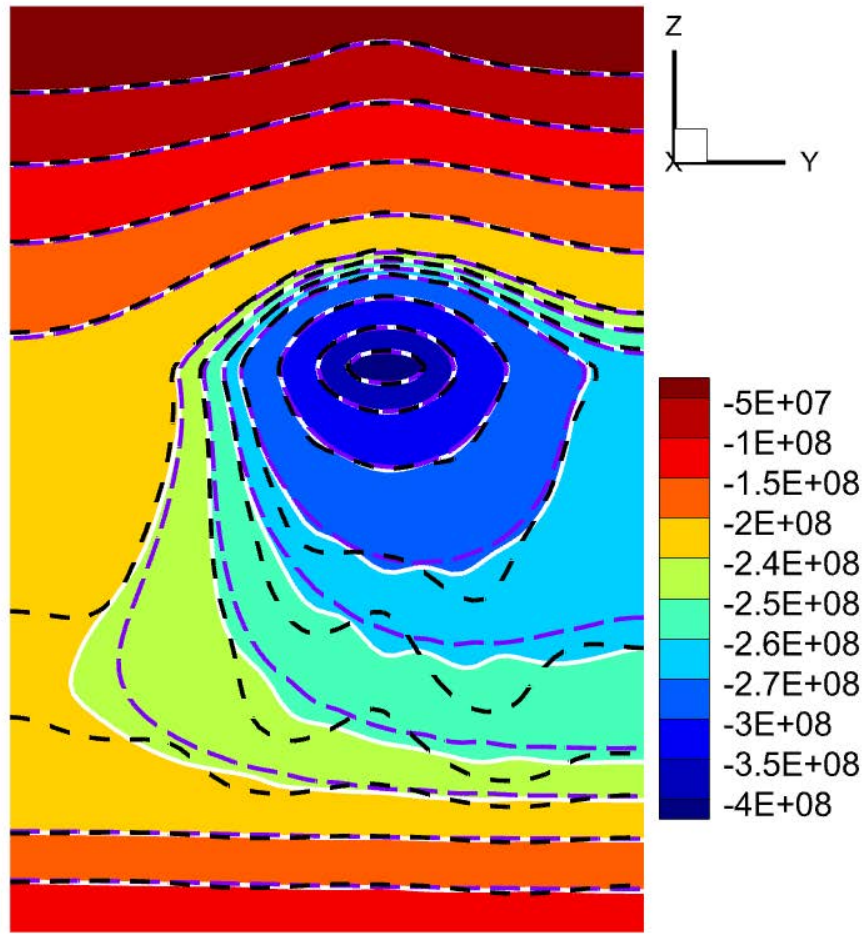}} ~~
\caption{(a) steady 3D temperature contour,  (c,d,e) quasi-2D temperature contour, heat flux along the $y$ and $z$ direction in the YZ slice in the bulk FinFET (\cref{quasi3dhot}) with different number of discrete solid angles, where $48^2$ discretized solid angles are used for colored background with white solid line,  $24^2$ discretized solid angles are used for black dashed line,  $96^2$ discretized solid angles are used for purple long dashed line. }
\label{DV_independence}
\end{figure}

Take the heat conduction in 3D bulk FinFET~(\cref{quasi3dhot}) as an example, we conduct an independent verification of the number of discrete solid angles.
Steady temperature contour or heat flux contour under `Continuous' heating source with different number of discrete solid angle is shown in~\cref{DV_independence}.
Numerical results show that $24^2$ discrete solid angles are not enough to accurately capture the non-diffusive heat conduction, and there is serious numerical jitter.
Temperature fields predicted by BTE with $48^2$ or $96^2$ discretized solid angles are almost the same in quasi-2D simulations.
Considering the huge computational amount of 3D simulation, we only use $48^2$ discrete points in order to take into account both accuracy and computational efficiency.

\section{Macroscopic heat diffusion equation}
\label{sec:diffusionsolver}

The macroscopic heat diffusion equation is
\begin{align}
C \frac{\partial T}{\partial t} - \nabla \cdot \left(  \kappa_{eff}(\bm{x}) \nabla T \right) =\dot{S},
\label{eq:effectiveFourier}
\end{align}
where effective thermal conductivity $\kappa_{eff}$ in the present simulation is a scalar rather than a tensor for a given spatial position $\bm{x}$.
Finite volume method is invoked and above diffusion equation in integral form over a control volume $i$ from time $t_n$ to $t_{n+1}=t_{n}+ \Delta t$ can be written as follows,
\begin{align}
&C_i  \frac{T_i^{n+1}- T_i^{n}}{\Delta  t} -\frac{1}{2} \sum_{j\in N(i)} S_{ij} \mathbf{n}_{ij}   \cdot  \left(  \kappa_{eff}(\bm{x}_{ij})  \nabla  T_{ij}^{n+1} + \kappa_{eff}(\bm{x}_{ij})  \nabla  T_{ij}^{n}   \right) \notag \\
&= \frac{ \dot{S}_i ^n +\dot{S}_i ^{n+1} }{2},  \\
&C_i  \frac{ \Delta T_i^{n} }{\Delta  t} -\frac{1}{2} \sum_{j\in N(i)} S_{ij} \mathbf{n}_{ij}   \cdot  \left( \kappa_{eff}(\bm{x}_{ij})  \nabla  (\Delta T_{ij}^{n})  \right) \notag \\
&= \frac{ \dot{S}_i ^n +\dot{S}_i ^{n+1} }{2} + \sum_{j\in N(i)} S_{ij} \mathbf{n}_{ij}   \cdot  \left( \kappa_{eff}(\bm{x}_{ij})  \nabla  T_{ij}^{n}   \right) ,
\label{eq:EFL}
\end{align}
where trapezoidal quadrature is used for the time integration of the diffusion and heat source terms, $\Delta T^n= T^{n+1}-T^{n}$, $V_i$ is the volume of the cell $i$, $N(i)$ is the sets of face neighbor cells of cell $i$, $ij$ is the interface between the cell $i$ and cell $j$, $S_{ij}$ is the area of the interface $ij$, $ \kappa_{eff}(\bm{x}_{ij}) $ is the effective thermal conductivity at the cell interface $\bm{x}_{ij}$, and $\mathbf{n}_{ij}$ is the normal of the interface $ij$ directing from the cell $i$ to the cell $j$. 
Above discretizations theoretically have second-order spatial and temporal accuracy.
The time step is set as $\Delta t=0.05 t_p= 0.125$ ps and Cartesian grids with uniform cell size $\Delta x=1$ nm is used.
Conjugate gradient method is used to solve this heat diffusion equation~\eqref{eq:EFL}.

To solve Eq.~\eqref{eq:EFL}, one of the key input parameter is the effective thermal conductivity at the interface between medium $1$ and $2$.
Let's take the quasi-1D heat conduction as an example and briefly introduce how to deal with interface thermal resistance efficiently for macroscopic diffusion equation.
Considering two discretized uniform cells adjacent to the interface between medium $1$ and $2$ with cell size $\Delta x$, as shown in~\cref{BTEdmm}(c), there are different temperature distributions near the interface.
The positions and temperatures of cell center are $(x_1,T_1)$ and $(x_2,T_2)$, respectively.
The positions and temperatures of the left and right limits of the interface are $(x_L,T_L)$ and $(x_R,T_R)$, respectively, where $\Delta x/2= x_L -x_1= x_2 -x_R$.
The heat flux at the left or right limit of the interface can be calculated based on the EFL,
\begin{align}
q_1 &= -\kappa_1 \frac{dT}{dx} \approx -\kappa_1 \frac{T_L -T_1}{x_L -x_1},  \\
q_2 &= -\kappa_2 \frac{dT}{dx} \approx -\kappa_2 \frac{T_R -T_2}{x_R -x_2}.  
\end{align}
The heat flux across the interface can be calculated according to the definition of interface thermal resistance,
\begin{align}
q_{f}= \frac{ T_L- T_R }{R_{eff}},
\end{align}
where $R_{eff} = \left( 4- 2  (  t_{12} +t_{21} )  \right)  /  \left( t_{12} C_1 v_1  \right) $~\cite{zeng_disparate_2014}.
Note that the heat flux at the left or right limit of the interface and the heat flux across the interface should be equal, i.e.,
\begin{align}
q_1=q_2=q_{f}.
\end{align}
Combined above four equations, we can obtain an approximated thermal conductivity $\kappa_{f}$ at the interface,
\begin{align}
q_{f}  &= -\kappa_{f}  \frac{T_1 -T_2}{x_1 -x_2}, \\
\kappa_{f} &=\frac{1}{ 0.5/\kappa_1 +0.5/ \kappa_2 + R_{eff}/ \Delta x  }.
\end{align}

The effective thermal conductivity in the macroscopic diffusion equation~\eqref{eq:effectiveFourier} is
\begin{align}
\kappa_{eff} (\bm{x}) &= \frac{1}{3} C v_g ^2  \tau_{eff} ,  \\
\tau_{eff} ^{-1} &= v_g/ \lambda  + \tau_{boundary} ^{-1},  \label{eq:EFLBC}
\end{align}
where the boundary scattering rate is $\tau_{boundary} ^{-1}= v_g / L_{eff} $, $L_{eff}$ is smallest characteristic length which depends on the spatial position $\bm{x}$.
Specific values of $L_{eff}$ in different spatial regions in this paper are list below.
In bulk or SOI FinFET, we set $L_{eff} = 8$ nm in the fin area and $L_{eff} = 10$ nm in the bottom silicon substrate areas.
In the silicon dioxide, the bulk thermal conductivity $1.4$ W$\cdot$ m$^{-1} \cdot $K$^{-1}$ is always used.

\bibliographystyle{elsarticle-num-names_clear}
\bibliography{phonon}

\begin{thebibliography}{52}
\expandafter\ifx\csname natexlab\endcsname\relax\def\natexlab#1{#1}\fi
\providecommand{\url}[1]{\texttt{#1}}
\providecommand{\href}[2]{#2}
\providecommand{\path}[1]{#1}
\providecommand{\DOIprefix}{doi:}
\providecommand{\ArXivprefix}{arXiv:}
\providecommand{\URLprefix}{URL: }
\providecommand{\Pubmedprefix}{pmid:}
\providecommand{\doi}[1]{\href{http://dx.doi.org/#1}{\path{#1}}}
\providecommand{\Pubmed}[1]{\href{pmid:#1}{\path{#1}}}
\providecommand{\bibinfo}[2]{#2}
\ifx\xfnm\relax \def\xfnm[#1]{\unskip,\space#1}\fi
\bibitem[{IEEE(2023)}]{IRDS2023}
\bibinfo{author}{IEEE}, \bibinfo{title}{International Roadmap for Devices and
  Systems ({IRDS™})}, \bibinfo{publisher}{IEEE}, \bibinfo{year}{2023}.
  \URLprefix \url{https://irds.ieee.org/editions/2023}.
\bibitem[{Das et~al.(2024)Das, Rajalekshmi, and James}]{Fin_GAAFET_2024_review}
\bibinfo{author}{R.~R. Das}, \bibinfo{author}{T.~R. Rajalekshmi},
  \bibinfo{author}{A.~James},
\newblock \bibinfo{title}{Finfet to gaa mbcfet: A review and insights},
\newblock \bibinfo{journal}{IEEE Access} \bibinfo{volume}{12}
  (\bibinfo{year}{2024}) \bibinfo{pages}{50556--50577}.
  \DOIprefix\doi{10.1109/ACCESS.2024.3384428}.
\bibitem[{Alam et~al.(2019)Alam, Mahajan, Chen, Ahn, Jiang, and
  Shin}]{heat_dissipations_cross_Scale_review_2019}
\bibinfo{author}{M.~A. Alam}, \bibinfo{author}{B.~K. Mahajan},
  \bibinfo{author}{Y.-P. Chen}, \bibinfo{author}{W.~Ahn},
  \bibinfo{author}{H.~Jiang}, \bibinfo{author}{S.~H. Shin},
\newblock \bibinfo{title}{A device-to-system perspective regarding self-heating
  enhanced hot carrier degradation in modern field-effect transistors: A
  topical review},
\newblock \bibinfo{journal}{IEEE Transactions on Electron Devices}
  \bibinfo{volume}{66} (\bibinfo{year}{2019}) \bibinfo{pages}{4556--4565}.
  \DOIprefix\doi{10.1109/TED.2019.2941445}.
\bibitem[{Pop and Goodson(2004)}]{pop2004transistors}
\bibinfo{author}{E.~Pop}, \bibinfo{author}{K.~Goodson},
\newblock \bibinfo{title}{Thermal phenomena in nanoscale transistors},
\newblock in: \bibinfo{booktitle}{The Ninth Intersociety Conference on Thermal
  and Thermomechanical Phenomena In Electronic Systems (IEEE Cat.
  No.04CH37543)}, volume~\bibinfo{volume}{1}, \bibinfo{year}{2004}, pp.
  \bibinfo{pages}{1--7 Vol.1}. \DOIprefix\doi{10.1109/ITHERM.2004.1319147}.
\bibitem[{Qu et~al.(2020)Qu, Lu, Li, Chen, Zhang, Li, Lee, and
  Zhao}]{experiments_2020_transient}
\bibinfo{author}{Y.~Qu}, \bibinfo{author}{J.~Lu}, \bibinfo{author}{J.~Li},
  \bibinfo{author}{Z.~Chen}, \bibinfo{author}{J.~Zhang},
  \bibinfo{author}{C.~Li}, \bibinfo{author}{S.-W. Lee},
  \bibinfo{author}{Y.~Zhao},
\newblock \bibinfo{title}{In-situ monitoring of self-heating effect in
  aggressively scaled finfets and its quantitative impact on hot carrier
  degradation under dynamic circuit operation},
\newblock in: \bibinfo{booktitle}{2020 IEEE International Reliability Physics
  Symposium (IRPS)}, \bibinfo{year}{2020}, pp. \bibinfo{pages}{1--6}.
  \DOIprefix\doi{10.1109/IRPS45951.2020.9129591}.
\bibitem[{Ran~CHENG(2024)}]{English_rancheng_low_temperature_FinFET2024}
\bibinfo{author}{Z.~W. J. Z. W. S. J. Z. Y. G.~H. Ran~CHENG, Bo~LI},
\newblock \bibinfo{title}{Low-temperature cmos technology for high-performance
  computing: development and challenges},
\newblock \bibinfo{journal}{SCIENTIA SINICA Informationis} \bibinfo{volume}{54}
  (\bibinfo{year}{2024}) \bibinfo{pages}{88--}. \URLprefix
  \url{http://www.sciengine.com/publisher/Science China Press/journal/SCIENTIA
  SINICA Informationis/54/1/10.1360/SSI-2023-0347}.
  \DOIprefix\doi{https://doi.org/10.1360/SSI-2023-0347}.
\bibitem[{Wang and Vafai(2024)}]{WANG2024121499}
\bibinfo{author}{C.~Wang}, \bibinfo{author}{K.~Vafai},
\newblock \bibinfo{title}{Heat transfer enhancement for 3d chip thermal
  simulation and prediction},
\newblock \bibinfo{journal}{Applied Thermal Engineering} \bibinfo{volume}{236}
  (\bibinfo{year}{2024}) \bibinfo{pages}{121499}. \URLprefix
  \url{https://www.sciencedirect.com/science/article/pii/S1359431123015284}.
  \DOIprefix\doi{https://doi.org/10.1016/j.applthermaleng.2023.121499}.
\bibitem[{Kim et~al.(2019)Kim, Lee, Park, and Seol}]{KIM2019114080}
\bibinfo{author}{C.~Kim}, \bibinfo{author}{M.~Lee}, \bibinfo{author}{J.~Park},
  \bibinfo{author}{J.~H. Seol},
\newblock \bibinfo{title}{Measurement and analysis of ballistic-diffusive
  phonon heat transport in a constrained silicon film},
\newblock \bibinfo{journal}{Applied Thermal Engineering} \bibinfo{volume}{160}
  (\bibinfo{year}{2019}) \bibinfo{pages}{114080}. \URLprefix
  \url{https://www.sciencedirect.com/science/article/pii/S1359431119307860}.
  \DOIprefix\doi{https://doi.org/10.1016/j.applthermaleng.2019.114080}.
\bibitem[{Ziabari et~al.(2018)Ziabari, Torres, Vermeersch, Xuan, Cartoix{\`a},
  Torell{\'o}, Bahk, Koh, Parsa, Ye, Alvarez, and Shakouri}]{ziabari2018a}
\bibinfo{author}{A.~Ziabari}, \bibinfo{author}{P.~Torres},
  \bibinfo{author}{B.~Vermeersch}, \bibinfo{author}{Y.~Xuan},
  \bibinfo{author}{X.~Cartoix{\`a}}, \bibinfo{author}{A.~Torell{\'o}},
  \bibinfo{author}{J.-H. Bahk}, \bibinfo{author}{Y.~R. Koh},
  \bibinfo{author}{M.~Parsa}, \bibinfo{author}{P.~D. Ye},
  \bibinfo{author}{F.~X. Alvarez}, \bibinfo{author}{A.~Shakouri},
\newblock \bibinfo{title}{Full-field thermal imaging of quasiballistic
  crosstalk reduction in nanoscale devices},
\newblock \bibinfo{journal}{Nat. Commun.} \bibinfo{volume}{9}
  (\bibinfo{year}{2018}) \bibinfo{pages}{255}. \URLprefix
  \url{https://www.nature.com/articles/s41467-017-02652-4}.
  \DOIprefix\doi{10.1038/s41467-017-02652-4}.
\bibitem[{Gu et~al.(2018)Gu, Wei, Yin, Li, and Yang}]{RevModPhys.90.041002}
\bibinfo{author}{X.~Gu}, \bibinfo{author}{Y.~Wei}, \bibinfo{author}{X.~Yin},
  \bibinfo{author}{B.~Li}, \bibinfo{author}{R.~Yang},
\newblock \bibinfo{title}{Colloquium: phononic thermal properties of
  two-dimensional materials},
\newblock \bibinfo{journal}{Rev. Mod. Phys.} \bibinfo{volume}{90}
  (\bibinfo{year}{2018}) \bibinfo{pages}{041002}. \URLprefix
  \url{https://link.aps.org/doi/10.1103/RevModPhys.90.041002}.
  \DOIprefix\doi{10.1103/RevModPhys.90.041002}.
\bibitem[{Zhang et~al.(2020)Zhang, Ouyang, Cheng, Chen, Li, and
  Zhang}]{ZHANG2020}
\bibinfo{author}{Z.~Zhang}, \bibinfo{author}{Y.~Ouyang},
  \bibinfo{author}{Y.~Cheng}, \bibinfo{author}{J.~Chen},
  \bibinfo{author}{N.~Li}, \bibinfo{author}{G.~Zhang},
\newblock \bibinfo{title}{Size-dependent phononic thermal transport in
  low-dimensional nanomaterials},
\newblock \bibinfo{journal}{Phys. Rep.} \bibinfo{volume}{860}
  (\bibinfo{year}{2020}) \bibinfo{pages}{1--26}. \URLprefix
  \url{https://www.sciencedirect.com/science/article/pii/S0370157320300922}.
  \DOIprefix\doi{10.1016/j.physrep.2020.03.001}.
\bibitem[{Stettler et~al.(2021)Stettler, Cea, Hasan, Jiang, Keys, Landon,
  Marepalli, Pantuso, and Weber}]{TCAD_application_intel_2021_review}
\bibinfo{author}{M.~A. Stettler}, \bibinfo{author}{S.~M. Cea},
  \bibinfo{author}{S.~Hasan}, \bibinfo{author}{L.~Jiang},
  \bibinfo{author}{P.~H. Keys}, \bibinfo{author}{C.~D. Landon},
  \bibinfo{author}{P.~Marepalli}, \bibinfo{author}{D.~Pantuso},
  \bibinfo{author}{C.~E. Weber},
\newblock \bibinfo{title}{Industrial {TCAD}: Modeling atoms to chips},
\newblock \bibinfo{journal}{IEEE Transactions on Electron Devices}
  \bibinfo{volume}{68} (\bibinfo{year}{2021}) \bibinfo{pages}{5350--5357}.
  \DOIprefix\doi{10.1109/TED.2021.3076976}.
\bibitem[{Pop(2010)}]{pop_energy_2010}
\bibinfo{author}{E.~Pop},
\newblock \bibinfo{title}{Energy dissipation and transport in nanoscale
  devices},
\newblock \bibinfo{journal}{Nano Res.} \bibinfo{volume}{3}
  (\bibinfo{year}{2010}) \bibinfo{pages}{147--169}. \URLprefix
  \url{https://doi.org/10.1007/s12274-010-1019-z}.
  \DOIprefix\doi{10.1007/s12274-010-1019-z}.
\bibitem[{Mishra et~al.(2024)Mishra, Vermeersch, Sankatali, Kukner, Mirabelli,
  Bufler, Brunion, Abdi, Oprins, Biswas, Zografos, Catthoor, Weckx, Hellings,
  Myers, and Ryckaert}]{2024_VLSI_thermal_IMEC}
\bibinfo{author}{S.~Mishra}, \bibinfo{author}{B.~Vermeersch},
  \bibinfo{author}{V.~Sankatali}, \bibinfo{author}{H.~Kukner},
  \bibinfo{author}{G.~Mirabelli}, \bibinfo{author}{F.~M. Bufler},
  \bibinfo{author}{M.~Brunion}, \bibinfo{author}{D.~Abdi},
  \bibinfo{author}{H.~Oprins}, \bibinfo{author}{D.~Biswas},
  \bibinfo{author}{O.~Zografos}, \bibinfo{author}{F.~Catthoor},
  \bibinfo{author}{P.~Weckx}, \bibinfo{author}{G.~Hellings},
  \bibinfo{author}{J.~Myers}, \bibinfo{author}{J.~Ryckaert},
\newblock \bibinfo{title}{Thermal considerations for block-level ppa assessment
  in angstrom era: A comparison study of nanosheet fets (a10) \& complementary
  fets (a5)},
\newblock in: \bibinfo{booktitle}{2024 IEEE Symposium on VLSI Technology and
  Circuits (VLSI Technology and Circuits)}, \bibinfo{year}{2024}, pp.
  \bibinfo{pages}{1--2}.
  \DOIprefix\doi{10.1109/VLSITechnologyandCir46783.2024.10631358}.
\bibitem[{Chang et~al.(2023)Chang, Oprins, Lofrano, Cherman, Vermeersch,
  Fortuny, Park, Tokei, and De~Wolf}]{BEOL2023_IMEC}
\bibinfo{author}{X.~Chang}, \bibinfo{author}{H.~Oprins},
  \bibinfo{author}{M.~Lofrano}, \bibinfo{author}{V.~Cherman},
  \bibinfo{author}{B.~Vermeersch}, \bibinfo{author}{J.~D. Fortuny},
  \bibinfo{author}{S.~Park}, \bibinfo{author}{Z.~Tokei},
  \bibinfo{author}{I.~De~Wolf},
\newblock \bibinfo{title}{Calibrated fast thermal calculation and experimental
  characterization of advanced {BEOL} stacks},
\newblock in: \bibinfo{booktitle}{2023 IEEE International Interconnect
  Technology Conference (IITC) and IEEE Materials for Advanced Metallization
  Conference (MAM)(IITC/MAM)}, \bibinfo{year}{2023}, pp. \bibinfo{pages}{1--3}.
  \DOIprefix\doi{10.1109/IITC/MAM57687.2023.10154768}.
\bibitem[{Zhang et~al.(2024)Zhang, Guo, Lian, and Shiomi}]{ZHANG2024123379}
\bibinfo{author}{C.~Zhang}, \bibinfo{author}{R.~Guo},
  \bibinfo{author}{M.~Lian}, \bibinfo{author}{J.~Shiomi},
\newblock \bibinfo{title}{Electron–phonon coupling and non-equilibrium
  thermal conduction in ultra-fast heating systems},
\newblock \bibinfo{journal}{Applied Thermal Engineering} \bibinfo{volume}{249}
  (\bibinfo{year}{2024}) \bibinfo{pages}{123379}. \URLprefix
  \url{https://www.sciencedirect.com/science/article/pii/S1359431124010470}.
  \DOIprefix\doi{https://doi.org/10.1016/j.applthermaleng.2024.123379}.
\bibitem[{Hao et~al.(2018)Hao, Zhao, Xiao, Wang, and Wang}]{IEEE_qinghao_2018}
\bibinfo{author}{Q.~Hao}, \bibinfo{author}{H.~Zhao}, \bibinfo{author}{Y.~Xiao},
  \bibinfo{author}{Q.~Wang}, \bibinfo{author}{X.~Wang},
\newblock \bibinfo{title}{Hybrid electrothermal simulation of a 3-d fin-shaped
  field-effect transistor based on gan nanowires},
\newblock \bibinfo{journal}{IEEE Transactions on Electron Devices}
  \bibinfo{volume}{65} (\bibinfo{year}{2018}) \bibinfo{pages}{921--927}.
  \DOIprefix\doi{10.1109/TED.2018.2791959}.
\bibitem[{Landon et~al.(2023)Landon, Jiang, Pantuso, Meric, Komeyli, Hicks, and
  Schroeder}]{intel_2023_GAAFET}
\bibinfo{author}{C.~Landon}, \bibinfo{author}{L.~Jiang},
  \bibinfo{author}{D.~Pantuso}, \bibinfo{author}{I.~Meric},
  \bibinfo{author}{K.~Komeyli}, \bibinfo{author}{J.~Hicks},
  \bibinfo{author}{D.~Schroeder},
\newblock \bibinfo{title}{Localized thermal effects in gate-all-around
  devices},
\newblock in: \bibinfo{booktitle}{2023 IEEE International Reliability Physics
  Symposium (IRPS)}, \bibinfo{year}{2023}, pp. \bibinfo{pages}{1--5}.
  \DOIprefix\doi{10.1109/IRPS48203.2023.10117903}.
\bibitem[{Zhang et~al.(2025)Zhang, Lou, and Liang}]{ZHANG2025126374}
\bibinfo{author}{C.~Zhang}, \bibinfo{author}{Q.~Lou},
  \bibinfo{author}{H.~Liang},
\newblock \bibinfo{title}{Synthetic iterative scheme for thermal applications
  in hotspot systems with large temperature variance},
\newblock \bibinfo{journal}{International Journal of Heat and Mass Transfer}
  \bibinfo{volume}{236} (\bibinfo{year}{2025}) \bibinfo{pages}{126374}.
  \URLprefix
  \url{https://www.sciencedirect.com/science/article/pii/S0017931024012031}.
  \DOIprefix\doi{https://doi.org/10.1016/j.ijheatmasstransfer.2024.126374}.
\bibitem[{Shen and Cao(2024)}]{shen2024_APL}
\bibinfo{author}{Y.~Shen}, \bibinfo{author}{B.~Cao},
\newblock \bibinfo{title}{Two-temperature principle for evaluating
  electrothermal performance of gan hemts},
\newblock \bibinfo{journal}{Applied Physics Letters} \bibinfo{volume}{124}
  (\bibinfo{year}{2024}) \bibinfo{pages}{042107}. \URLprefix
  \url{https://doi.org/10.1063/5.0189262}. \DOIprefix\doi{10.1063/5.0189262}.
\bibitem[{Sendra et~al.(2021)Sendra, Beardo, Torres, Bafaluy, Alvarez, and
  Camacho}]{PhysRevB.103.L140301}
\bibinfo{author}{L.~Sendra}, \bibinfo{author}{A.~Beardo},
  \bibinfo{author}{P.~Torres}, \bibinfo{author}{J.~Bafaluy},
  \bibinfo{author}{F.~X. Alvarez}, \bibinfo{author}{J.~Camacho},
\newblock \bibinfo{title}{Derivation of a hydrodynamic heat equation from the
  phonon boltzmann equation for general semiconductors},
\newblock \bibinfo{journal}{Phys. Rev. B} \bibinfo{volume}{103}
  (\bibinfo{year}{2021}) \bibinfo{pages}{L140301}. \URLprefix
  \url{https://link.aps.org/doi/10.1103/PhysRevB.103.L140301}.
  \DOIprefix\doi{10.1103/PhysRevB.103.L140301}.
\bibitem[{Joseph and Preziosi(1989)}]{RevModPhysJoseph89}
\bibinfo{author}{D.~D. Joseph}, \bibinfo{author}{L.~Preziosi},
\newblock \bibinfo{title}{Heat waves},
\newblock \bibinfo{journal}{Rev. Mod. Phys.} \bibinfo{volume}{61}
  (\bibinfo{year}{1989}) \bibinfo{pages}{41--73}. \URLprefix
  \url{https://link.aps.org/doi/10.1103/RevModPhys.61.41}.
  \DOIprefix\doi{10.1103/RevModPhys.61.41}.
\bibitem[{Kovács(2024)}]{KOVACS20241}
\bibinfo{author}{R.~Kovács},
\newblock \bibinfo{title}{Heat equations beyond fourier: From heat waves to
  thermal metamaterials},
\newblock \bibinfo{journal}{Phys. Rep.} \bibinfo{volume}{1048}
  (\bibinfo{year}{2024}) \bibinfo{pages}{1--75}. \URLprefix
  \url{https://www.sciencedirect.com/science/article/pii/S0370157323003770}.
  \DOIprefix\doi{https://doi.org/10.1016/j.physrep.2023.11.001}.
\bibitem[{Rezgui et~al.(2021)Rezgui, Nasri, Ali, and
  Guizani}]{hydrodynamicsHoussem2021}
\bibinfo{author}{H.~Rezgui}, \bibinfo{author}{F.~Nasri},
  \bibinfo{author}{A.~B.~H. Ali}, \bibinfo{author}{A.~A. Guizani},
\newblock \bibinfo{title}{Analysis of the ultrafast transient heat transport in
  sub 7-nm {SOI FinFETs} technology nodes using phonon hydrodynamic equation},
\newblock \bibinfo{journal}{IEEE Transactions on Electron Devices}
  \bibinfo{volume}{68} (\bibinfo{year}{2021}) \bibinfo{pages}{10--16}.
  \DOIprefix\doi{10.1109/TED.2020.3039200}.
\bibitem[{Zhang et~al.(2023)Zhang, Huberman, Song, Zhao, Chen, and
  Wu}]{ZHANG2023124715}
\bibinfo{author}{C.~Zhang}, \bibinfo{author}{S.~Huberman},
  \bibinfo{author}{X.~Song}, \bibinfo{author}{J.~Zhao},
  \bibinfo{author}{S.~Chen}, \bibinfo{author}{L.~Wu},
\newblock \bibinfo{title}{Acceleration strategy of source iteration method for
  the stationary phonon boltzmann transport equation},
\newblock \bibinfo{journal}{Int. J. Heat Mass Transfer} \bibinfo{volume}{217}
  (\bibinfo{year}{2023}) \bibinfo{pages}{124715}. \URLprefix
  \url{https://www.sciencedirect.com/science/article/pii/S0017931023008608}.
  \DOIprefix\doi{https://doi.org/10.1016/j.ijheatmasstransfer.2023.124715}.
\bibitem[{Mazumder(2022)}]{mazumder_boltzmann_2022}
\bibinfo{author}{S.~Mazumder},
\newblock \bibinfo{title}{{Boltzmann} transport equation based modeling of
  phonon heat conduction: {Progress} and challenges},
\newblock \bibinfo{journal}{Annual Review of Heat Transfer}
  \bibinfo{volume}{24} (\bibinfo{year}{2022}) \bibinfo{pages}{71--130}.
  \URLprefix
  \url{https://www.dl.begellhouse.com/references/5756967540dd1b03,3ae07302147f45b7,09643dee3a7e400e.html}.
  \DOIprefix\doi{10.1615/AnnualRevHeatTransfer.2022041316}.
\bibitem[{Barry et~al.(2022)Barry, Kumar, and Kumar}]{barry2022boltzmann}
\bibinfo{author}{M.~C. Barry}, \bibinfo{author}{N.~Kumar},
  \bibinfo{author}{S.~Kumar},
\newblock \bibinfo{title}{{Boltzmann} transport equation for thermal transport
  in electronic materials and devices},
\newblock \bibinfo{journal}{Annual Review of Heat Transfer}
  \bibinfo{volume}{24} (\bibinfo{year}{2022}) \bibinfo{pages}{131--172}.
  \URLprefix
  \url{https://dl.begellhouse.com/references/5756967540dd1b03,3ae07302147f45b7,43e971b72ed09c31.html}.
  \DOIprefix\doi{10.1615/AnnualRevHeatTransfer.v24.50}.
\bibitem[{Hu et~al.(2024)Hu, Shen, and Bao}]{HU2024huabao}
\bibinfo{author}{Y.~Hu}, \bibinfo{author}{Y.~Shen}, \bibinfo{author}{H.~Bao},
\newblock \bibinfo{title}{Ultra-efficient and parameter-free computation of
  submicron thermal transport with phonon boltzmann transport equation},
\newblock \bibinfo{journal}{Fundamental Research} \bibinfo{volume}{4}
  (\bibinfo{year}{2024}) \bibinfo{pages}{907--915}. \URLprefix
  \url{https://www.sciencedirect.com/science/article/pii/S2667325822002758}.
  \DOIprefix\doi{https://doi.org/10.1016/j.fmre.2022.06.007}.
\bibitem[{Pham et~al.(2018)Pham, Jin, Lu, Park, Choi, Pourghaderi, Kim, Kwon,
  and Kim}]{coupled_samsung2018}
\bibinfo{author}{A.-T. Pham}, \bibinfo{author}{S.~Jin},
  \bibinfo{author}{Y.~Lu}, \bibinfo{author}{H.-H. Park},
  \bibinfo{author}{W.~Choi}, \bibinfo{author}{M.~A. Pourghaderi},
  \bibinfo{author}{J.~Kim}, \bibinfo{author}{U.~Kwon},
  \bibinfo{author}{D.~Kim},
\newblock \bibinfo{title}{Simulations of self-heating effects in sige pfinfets
  based on self-consistent solution of carrier/phonon bte coupled system},
\newblock in: \bibinfo{booktitle}{2018 International Conference on Simulation
  of Semiconductor Processes and Devices (SISPAD)}, \bibinfo{year}{2018}, pp.
  \bibinfo{pages}{145--148}. \DOIprefix\doi{10.1109/SISPAD.2018.8551670}.
\bibitem[{Lofrano et~al.(2023)Lofrano, Oprins, Chang, Vermeersch, Pedreira,
  Lesniewska, Cherman, Ciofi, Croes, Park, and Tokei}]{2023IEEEthermal_IMEC}
\bibinfo{author}{M.~Lofrano}, \bibinfo{author}{H.~Oprins},
  \bibinfo{author}{X.~Chang}, \bibinfo{author}{B.~Vermeersch},
  \bibinfo{author}{O.~V. Pedreira}, \bibinfo{author}{A.~Lesniewska},
  \bibinfo{author}{V.~Cherman}, \bibinfo{author}{I.~Ciofi},
  \bibinfo{author}{K.~Croes}, \bibinfo{author}{S.~Park},
  \bibinfo{author}{Z.~Tokei},
\newblock \bibinfo{title}{Towards accurate temperature prediction in {BEOL} for
  reliability assessment (invited)},
\newblock in: \bibinfo{booktitle}{2023 IEEE International Reliability Physics
  Symposium (IRPS)}, \bibinfo{year}{2023}, pp. \bibinfo{pages}{1--7}.
  \DOIprefix\doi{10.1109/IRPS48203.2023.10117701}.
\bibitem[{Xu et~al.(2023)Xu, Hu, and Bao}]{PhysRevApplied.19.014007}
\bibinfo{author}{J.~Xu}, \bibinfo{author}{Y.~Hu}, \bibinfo{author}{H.~Bao},
\newblock \bibinfo{title}{Quantitative analysis of nonequilibrium phonon
  transport near a nanoscale hotspot},
\newblock \bibinfo{journal}{Phys. Rev. Appl.} \bibinfo{volume}{19}
  (\bibinfo{year}{2023}) \bibinfo{pages}{014007}. \URLprefix
  \url{https://link.aps.org/doi/10.1103/PhysRevApplied.19.014007}.
  \DOIprefix\doi{10.1103/PhysRevApplied.19.014007}.
\bibitem[{Sheng et~al.(2024)Sheng, Wang, Hu, Xu, Ji, and
  Bao}]{baohua_IEEE_2024}
\bibinfo{author}{Y.~Sheng}, \bibinfo{author}{S.~Wang}, \bibinfo{author}{Y.~Hu},
  \bibinfo{author}{J.~Xu}, \bibinfo{author}{Z.~Ji}, \bibinfo{author}{H.~Bao},
\newblock \bibinfo{title}{Integrating first-principles-based {Non-Fourier}
  thermal analysis into nanoscale device simulation},
\newblock \bibinfo{journal}{IEEE Transactions on Electron Devices}
  \bibinfo{volume}{71} (\bibinfo{year}{2024}) \bibinfo{pages}{1769--1775}.
  \DOIprefix\doi{10.1109/TED.2024.3357440}.
\bibitem[{Adisusilo et~al.(2014)Adisusilo, Kukita, and
  Kamakura}]{3DFINFET_2014_mc}
\bibinfo{author}{I.~N. Adisusilo}, \bibinfo{author}{K.~Kukita},
  \bibinfo{author}{Y.~Kamakura},
\newblock \bibinfo{title}{Analysis of heat conduction property in {FinFETs}
  using phonon {Monte Carlo} simulation},
\newblock in: \bibinfo{booktitle}{2014 International Conference on Simulation
  of Semiconductor Processes and Devices (SISPAD)}, \bibinfo{year}{2014}, pp.
  \bibinfo{pages}{17--20}. \DOIprefix\doi{10.1109/SISPAD.2014.6931552}.
\bibitem[{Mukhopadhyay et~al.(2018)Mukhopadhyay, Kundu, Lee, Hsieh, Huang,
  Horng, Chen, Lee, Tsai, Lin, Lu, and He}]{TSMC_2018_self_heating}
\bibinfo{author}{S.~Mukhopadhyay}, \bibinfo{author}{A.~Kundu},
  \bibinfo{author}{Y.~Lee}, \bibinfo{author}{H.~D. Hsieh},
  \bibinfo{author}{D.~Huang}, \bibinfo{author}{J.~Horng},
  \bibinfo{author}{T.~Chen}, \bibinfo{author}{J.~Lee},
  \bibinfo{author}{Y.~Tsai}, \bibinfo{author}{C.~Lin}, \bibinfo{author}{R.~Lu},
  \bibinfo{author}{J.~He},
\newblock \bibinfo{title}{An unique methodology to estimate the thermal time
  constant and dynamic self heating impact for accurate reliability evaluation
  in advanced {FinFET} technologies},
\newblock in: \bibinfo{booktitle}{2018 IEEE International Electron Devices
  Meeting (IEDM)}, \bibinfo{year}{2018}, pp. \bibinfo{pages}{17.4.1--17.4.4}.
  \DOIprefix\doi{10.1109/IEDM.2018.8614479}.
\bibitem[{Zhang and Wu(2022)}]{PhysRevE.106.014111}
\bibinfo{author}{C.~Zhang}, \bibinfo{author}{L.~Wu},
\newblock \bibinfo{title}{Nonmonotonic heat dissipation phenomenon in
  close-packed hotspot systems},
\newblock \bibinfo{journal}{Phys. Rev. E} \bibinfo{volume}{106}
  (\bibinfo{year}{2022}) \bibinfo{pages}{014111}. \URLprefix
  \url{https://link.aps.org/doi/10.1103/PhysRevE.106.014111}.
  \DOIprefix\doi{10.1103/PhysRevE.106.014111}.
\bibitem[{Guo and Xu(2016)}]{GuoZl16DUGKS}
\bibinfo{author}{Z.~Guo}, \bibinfo{author}{K.~Xu},
\newblock \bibinfo{title}{Discrete unified gas kinetic scheme for multiscale
  heat transfer based on the phonon {B}oltzmann transport equation},
\newblock \bibinfo{journal}{Int. J. Heat Mass Transfer} \bibinfo{volume}{102}
  (\bibinfo{year}{2016}) \bibinfo{pages}{944 -- 958}. \URLprefix
  \url{http://www.sciencedirect.com/science/article/pii/S0017931016306731}.
  \DOIprefix\doi{10.1016/j.ijheatmasstransfer.2016.06.088}.
\bibitem[{Tang and Cao(2023)}]{TANG2023123497}
\bibinfo{author}{D.-S. Tang}, \bibinfo{author}{B.-Y. Cao},
\newblock \bibinfo{title}{Phonon thermal transport and its tunability in {GaN}
  for near-junction thermal management of electronics: A review},
\newblock \bibinfo{journal}{Int. J. Heat Mass Transfer} \bibinfo{volume}{200}
  (\bibinfo{year}{2023}) \bibinfo{pages}{123497}. \URLprefix
  \url{https://www.sciencedirect.com/science/article/pii/S0017931022009668}.
  \DOIprefix\doi{https://doi.org/10.1016/j.ijheatmasstransfer.2022.123497}.
\bibitem[{Murthy et~al.(2005)Murthy, Narumanchi, Pascual-Gutierrez, Wang, Ni,
  and Mathur}]{MurthyJY05Review}
\bibinfo{author}{J.~Y. Murthy}, \bibinfo{author}{S.~V.~J. Narumanchi},
  \bibinfo{author}{J.~A. Pascual-Gutierrez}, \bibinfo{author}{T.~Wang},
  \bibinfo{author}{C.~Ni}, \bibinfo{author}{S.~R. Mathur},
\newblock \bibinfo{title}{Review of multiscale simulation in submicron heat
  transfer},
\newblock \bibinfo{journal}{Int. J. Multiscale Computat. Eng.}
  \bibinfo{volume}{3} (\bibinfo{year}{2005}) \bibinfo{pages}{5--32}. \URLprefix
  \url{http://dl.begellhouse.com/journals/61fd1b191cf7e96f,69f10ca36a816eb7,25fd09426d0aaf45.html}.
  \DOIprefix\doi{10.1615/IntJMultCompEng.v3.i1.20}.
\bibitem[{Warzoha et~al.(2021)Warzoha, Wilson, Donovan, Donmezer, Giri,
  Hopkins, Choi, Pahinkar, Shi, Graham, Tian, and
  Ruppalt}]{warzoha_applications_2021}
\bibinfo{author}{R.~J. Warzoha}, \bibinfo{author}{A.~A. Wilson},
  \bibinfo{author}{B.~F. Donovan}, \bibinfo{author}{N.~Donmezer},
  \bibinfo{author}{A.~Giri}, \bibinfo{author}{P.~E. Hopkins},
  \bibinfo{author}{S.~Choi}, \bibinfo{author}{D.~Pahinkar},
  \bibinfo{author}{J.~Shi}, \bibinfo{author}{S.~Graham},
  \bibinfo{author}{Z.~Tian}, \bibinfo{author}{L.~Ruppalt},
\newblock \bibinfo{title}{Applications and impacts of nanoscale thermal
  transport in electronics packaging},
\newblock \bibinfo{journal}{J Electron. Packaging} \bibinfo{volume}{143}
  (\bibinfo{year}{2021}) \bibinfo{pages}{020804}. \URLprefix
  \url{https://doi.org/10.1115/1.4049293}. \DOIprefix\doi{10.1115/1.4049293}.
\bibitem[{Goodson and Flik(1992)}]{Silicon-on-Insulator_1992_IEEE}
\bibinfo{author}{K.~Goodson}, \bibinfo{author}{M.~Flik},
\newblock \bibinfo{title}{Effect of microscale thermal conduction on the
  packing limit of silicon-on-insulator electronic devices},
\newblock \bibinfo{journal}{IEEE Transactions on Components, Hybrids, and
  Manufacturing Technology} \bibinfo{volume}{15} (\bibinfo{year}{1992})
  \bibinfo{pages}{715--722}. \DOIprefix\doi{10.1109/33.180035}.
\bibitem[{Yang et~al.(2005)Yang, Chen, Laroche, and Taur}]{YangRg05BDE}
\bibinfo{author}{R.~Yang}, \bibinfo{author}{G.~Chen},
  \bibinfo{author}{M.~Laroche}, \bibinfo{author}{Y.~Taur},
\newblock \bibinfo{title}{Simulation of nanoscale multidimensional transient
  heat conduction problems using ballistic-diffusive equations and phonon
  {B}oltzmann equation},
\newblock \bibinfo{journal}{J. Heat Transfer} \bibinfo{volume}{127}
  (\bibinfo{year}{2005}) \bibinfo{pages}{298--306}. \URLprefix
  \url{http://dx.doi.org/10.1115/1.1857941}. \DOIprefix\doi{10.1115/1.1857941}.
\bibitem[{Nasri et~al.(2015)Nasri, {Ben Aissa}, and Belmabrouk}]{NASRI2015206}
\bibinfo{author}{F.~Nasri}, \bibinfo{author}{M.~{Ben Aissa}},
  \bibinfo{author}{H.~Belmabrouk},
\newblock \bibinfo{title}{Microscale thermal conduction based on
  cattaneo-vernotte model in silicon on insulator and double gate mosfets},
\newblock \bibinfo{journal}{Applied Thermal Engineering} \bibinfo{volume}{76}
  (\bibinfo{year}{2015}) \bibinfo{pages}{206--211}. \URLprefix
  \url{https://www.sciencedirect.com/science/article/pii/S1359431114010564}.
  \DOIprefix\doi{https://doi.org/10.1016/j.applthermaleng.2014.11.038}.
\bibitem[{Zhang and Guo(2019)}]{zhang_discrete_2019}
\bibinfo{author}{C.~Zhang}, \bibinfo{author}{Z.~Guo},
\newblock \bibinfo{title}{Discrete unified gas kinetic scheme for multiscale
  heat transfer with arbitrary temperature difference},
\newblock \bibinfo{journal}{Int. J. Heat Mass Transfer} \bibinfo{volume}{134}
  (\bibinfo{year}{2019}) \bibinfo{pages}{1127--1136}. \URLprefix
  \url{http://www.sciencedirect.com/science/article/pii/S0017931018353031}.
  \DOIprefix\doi{10.1016/j.ijheatmasstransfer.2019.02.056}.
\bibitem[{Tian et~al.(2024)Tian, Wu, Hu, Ma, and Zhang}]{dengke_2024_APL}
\bibinfo{author}{S.~Tian}, \bibinfo{author}{T.~Wu}, \bibinfo{author}{S.~Hu},
  \bibinfo{author}{D.~Ma}, \bibinfo{author}{L.~Zhang},
\newblock \bibinfo{title}{{Boosting phonon transport across AlN/SiC interface
  by fast annealing amorphous layers}},
\newblock \bibinfo{journal}{Applied Physics Letters} \bibinfo{volume}{124}
  (\bibinfo{year}{2024}) \bibinfo{pages}{042202}. \URLprefix
  \url{https://doi.org/10.1063/5.0187793}. \DOIprefix\doi{10.1063/5.0187793}.
\bibitem[{Chen et~al.(2022)Chen, Xu, Zhou, and Li}]{RevModPhys.94.025002}
\bibinfo{author}{J.~Chen}, \bibinfo{author}{X.~Xu}, \bibinfo{author}{J.~Zhou},
  \bibinfo{author}{B.~Li},
\newblock \bibinfo{title}{Interfacial thermal resistance: Past, present, and
  future},
\newblock \bibinfo{journal}{Rev. Mod. Phys.} \bibinfo{volume}{94}
  (\bibinfo{year}{2022}) \bibinfo{pages}{025002}. \URLprefix
  \url{https://link.aps.org/doi/10.1103/RevModPhys.94.025002}.
  \DOIprefix\doi{10.1103/RevModPhys.94.025002}.
\bibitem[{Hao et~al.(2017)Hao, Zhao, and Xiao}]{JAP_qinghao_2017}
\bibinfo{author}{Q.~Hao}, \bibinfo{author}{H.~Zhao}, \bibinfo{author}{Y.~Xiao},
\newblock \bibinfo{title}{{A hybrid simulation technique for electrothermal
  studies of two-dimensional {GaN-on-SiC} high electron mobility transistors}},
\newblock \bibinfo{journal}{J. Appl. Phys.} \bibinfo{volume}{121}
  (\bibinfo{year}{2017}) \bibinfo{pages}{204501}. \URLprefix
  \url{https://doi.org/10.1063/1.4983761}. \DOIprefix\doi{10.1063/1.4983761}.
\bibitem[{Medlar and Hensel(2022)}]{3DFINFETtransient}
\bibinfo{author}{M.~P. Medlar}, \bibinfo{author}{E.~C. Hensel},
\newblock \bibinfo{title}{{Transient Three-Dimensional Thermal Simulation of a
  Fin Field-Effect Transistor With Electron–Phonon Heat Generation, Three
  Phonon Scattering, and Drift With Periodic Switching}},
\newblock \bibinfo{journal}{ASME Journal of Heat and Mass Transfer}
  \bibinfo{volume}{145} (\bibinfo{year}{2022}) \bibinfo{pages}{022501}.
  \URLprefix \url{https://doi.org/10.1115/1.4056002}.
  \DOIprefix\doi{10.1115/1.4056002}.
\bibitem[{Honarvar et~al.(2021)Honarvar, Knobloch, Frazer, Abad, McBennett,
  Hussein, Kapteyn, Murnane, and Hernandez-Charpak}]{honarvar_directional_2021}
\bibinfo{author}{H.~Honarvar}, \bibinfo{author}{J.~L. Knobloch},
  \bibinfo{author}{T.~D. Frazer}, \bibinfo{author}{B.~Abad},
  \bibinfo{author}{B.~McBennett}, \bibinfo{author}{M.~I. Hussein},
  \bibinfo{author}{H.~C. Kapteyn}, \bibinfo{author}{M.~M. Murnane},
  \bibinfo{author}{J.~N. Hernandez-Charpak},
\newblock \bibinfo{title}{Directional thermal channeling: {A} phenomenon
  triggered by tight packing of heat sources},
\newblock \bibinfo{journal}{Proceedings of the National Academy of Sciences}
  \bibinfo{volume}{118} (\bibinfo{year}{2021}) \bibinfo{pages}{e2109056118}.
  \URLprefix \url{https://www.pnas.org/doi/abs/10.1073/pnas.2109056118}.
  \DOIprefix\doi{10.1073/pnas.2109056118}.
\bibitem[{Chen et~al.(2018)Chen, Hua, Zhang, Ravichandran, and
  Minnich}]{PhysRevApplied.10.054068}
\bibinfo{author}{X.~Chen}, \bibinfo{author}{C.~Hua},
  \bibinfo{author}{H.~Zhang}, \bibinfo{author}{N.~K. Ravichandran},
  \bibinfo{author}{A.~J. Minnich},
\newblock \bibinfo{title}{Quasiballistic thermal transport from nanoscale
  heaters and the role of the spatial frequency},
\newblock \bibinfo{journal}{Phys. Rev. Applied} \bibinfo{volume}{10}
  (\bibinfo{year}{2018}) \bibinfo{pages}{054068}. \URLprefix
  \url{https://link.aps.org/doi/10.1103/PhysRevApplied.10.054068}.
  \DOIprefix\doi{10.1103/PhysRevApplied.10.054068}.
\bibitem[{Chhabria and Sapatnekar(2019)}]{heat_chip_2019}
\bibinfo{author}{V.~A. Chhabria}, \bibinfo{author}{S.~S. Sapatnekar},
\newblock \bibinfo{title}{Impact of self-heating on performance and reliability
  in finfet and gaafet designs},
\newblock in: \bibinfo{booktitle}{20th International Symposium on Quality
  Electronic Design (ISQED)}, \bibinfo{year}{2019}, pp.
  \bibinfo{pages}{235--240}. \DOIprefix\doi{10.1109/ISQED.2019.8697786}.
\bibitem[{Guo and Xu(2021)}]{guo_progress_DUGKS}
\bibinfo{author}{Z.~Guo}, \bibinfo{author}{K.~Xu},
\newblock \bibinfo{title}{Progress of discrete unified gas-kinetic scheme for
  multiscale flows},
\newblock \bibinfo{journal}{Adva. Aerodyn.} \bibinfo{volume}{3}
  (\bibinfo{year}{2021}) \bibinfo{pages}{6}. \URLprefix
  \url{https://doi.org/10.1186/s42774-020-00058-3}.
  \DOIprefix\doi{10.1186/s42774-020-00058-3}.
\bibitem[{Zeng and Chen(2014)}]{zeng_disparate_2014}
\bibinfo{author}{L.~Zeng}, \bibinfo{author}{G.~Chen},
\newblock \bibinfo{title}{Disparate quasiballistic heat conduction regimes from
  periodic heat sources on a substrate},
\newblock \bibinfo{journal}{J. Appl. Phys.} \bibinfo{volume}{116}
  (\bibinfo{year}{2014}) \bibinfo{pages}{064307}. \URLprefix
  \url{https://aip.scitation.org/doi/10.1063/1.4893299}.
  \DOIprefix\doi{10.1063/1.4893299}.

\end{thebibliography}
\end{document}